\documentclass[manuscript,screen]{acmart}
\usepackage{graphicx}

\usepackage{lineno}

\usepackage{adjustbox}
\usepackage{listings}
\usepackage{xcolor}
\usepackage{algorithm2e}
\usepackage{pifont}

\definecolor{lightgray}{rgb}{9, 9, .9}
\definecolor{darkgray}{rgb}{.4, .4, .4}
\definecolor{purple}{rgb}{0.65, 0.12, 0.82}
\definecolor{gray}{rgb}{0.4,0.4,0.4}
\definecolor{line-numbers}{rgb}{0.4,0.4,0.4}
\definecolor{tags}{rgb}{1, 0, 0}
\definecolor{darkblue}{rgb}{0.0,0.0,0.6}
\definecolor{cyan}{rgb}{0.0,0.6,0.6}
\definecolor{highlight}{HTML}{ffffff}

\usepackage{soul}

\newcommand{\hlc}[2][yellow]{{%
    \colorlet{foo}{#1}%
    \sethlcolor{foo}\hl{#2}}%
}

\lstdefinelanguage{Groovy}
{
  sensitive=true,%
    morecomment=[l]//,%
  morecomment=[s]{/}{/},%
  morestring=[b]",%
  morestring=[b]',%
  stringstyle=\color{black},
  identifierstyle=\color{darkblue},
  keywordstyle=\color{cyan},
  morekeywords={abstract,any,as,boolean,break,byte,case,catch,char,
  class, const,continue,def,default,do,double,else,extends,false,final,finally, float,for,goto,if,implements,import,instanceof,in,int,interface,label, long,native,new,null,package,private,protected,public,return,short, static,strictfp,super,switch,synchronized,this,throw,throws,transient, true,try,void,volatile,while,with}
}
\usepackage[T1]{fontenc}
\usepackage{listings,beramono}
\usepackage{tikz}
\makeatletter
\newenvironment{btHighlight}[1][]
{\begingroup\tikzset{bt@Highlight@par/.style={#1}}\begin{lrbox}{\@tempboxa}}
{\end{lrbox}\bt@HL@box[bt@Highlight@par]{\@tempboxa}\endgroup}

\newcommand\btHL[1][]{%
  \begin{btHighlight}[#1]\bgroup\aftergroup\bt@HL@endenv%
}
\def\bt@HL@endenv{%
  \end{btHighlight}%
  \egroup
}
\newcommand{\bt@HL@box}[2][]{%
  \tikz[#1]{%
    \pgfpathrectangle{\pgfpoint{1pt}{0pt}}{\pgfpoint{\wd #2}{\ht #2}}%
    \pgfusepath{use as bounding box}%
    \node[anchor=base west, fill=orange!30,outer sep=0pt,inner xsep=1pt, inner ysep=0pt, rounded corners=3pt, minimum height=\ht\strutbox+1pt,#1]{\raisebox{1pt}{\strut}\strut\usebox{#2}};
  }%
}
\makeatother

\lstdefinestyle{Groovy}{
    language={Groovy}, 
    moredelim=**[is][\btHL]{`}{`},
    moredelim=**[is][{\btHL[fill=green!30]}]{*}{*},
    moredelim=**[is][{\btHL[fill=cyan!30]}]{~}{~},
    extendedchars=true,
}

\usepackage{qtree}
\usepackage{multicol}        
\usepackage{makecell}
\usepackage{url}   
   
\begin{document}

\title{An Automated Approach for Privacy Leakage
Identification in IoT Apps}

\author{Bara' Nazzal}
\affiliation{%
  \institution{Ryerson University}
  \city{Toronto}
  \country{Canada}}
\email{bara.nazzal@ryerson.ca}

\author{Manar Alalfi}
\affiliation{%
  \institution{Ryerson University}
  \city{Toronto}
  \country{Canada}}
\email{manar.alalfi@ryerson.ca}

\renewcommand{\shortauthors}{Bara' Nazzal, Manar H. Alalfi}

\begin{abstract}
This paper presents a fully automated static analysis approach and a tool, Taint-Things, for the identification of tainted flows in SmartThings IoT apps. Taint-Things accurately identifies all tainted flows reported by one of the state-of-the-art tools with at least 4 times improved performance. Our approach reports potential vulnerable tainted flows in a form of a concise security slice, where the relevant parts of the code are given with the lines affecting the sensitive information, which could provide security auditors with an effective and precise tool to pinpoint security issues in SmartThings apps under test. We also present and test ways to add precision to Taint-Things by adding extra sensitivities; we provide different approaches for flow, path and context sensitive analyses through modules that can be added to Taint-Things. We present experiments to evaluate Taint-Things by running it on a SmartThings app dataset as well as testing for precision and recall on a set generated by a mutation framework to see how much coverage is achieved without adding false positives.  This shows an improvement in performance both in terms of speed up to 4 folds, as well as improving the precision avoiding false positives by providing a higher level of flow and path sensitivity analysis in comparison with one of state of the art tools. 
\end{abstract}



\maketitle



\section{Introduction}\label{ch:Introduction}
  Today, more devices such as everyday utilities, home appliances, cars and other items are being embedded with software and are getting connected through the internet, giving rise to the concept of Internet of Things (IoT). While this technology brings with it lots of advantages, it also opens the door for many vulnerabilities, making the study of the security aspects of this technology very crucial.
  
  One of the main concerns in the field of IoT is the potential risk of sensitive data leaking. And with the increasing popularity of the technology, tackling this issue becomes more necessary. By their nature, IoT apps and devices communicate through the internet, messages and notifications, but with bad coding practices this could end up posing a serious risk of exposing the users' private information. Furthermore, malicious applications could be specifically designed to hide their malicious behavior through undeclared breaches.

  In a report by the Open Web Application Security Project (OWASP) that listed top 10 IoT vulnerabilities for the year 2018 \cite{OWASP10}, it included the risk of insufficient privacy protection as number 6. This indicates the risk of apps using sensitive information in a non secure manner or without permission. With the vulnerability of insecure network services being number 2 on the list, this increases the importance of providing a way to track private information within the app and detecting whether the information can be sent over networks, whether as messages, notifications or through the internet, which might not be secure.
  \begin{table}
\centering
\caption{OWASP IoT Top 10 Vulnerabilities \cite{OWASP10}}
\label{tabl:OWASP}
\begin{tabular}{|l|}
\hline
I1  Weak, Guessable, or Hardcoded Passwords \\ \hline
I2  Insecure Network Services               \\ \hline
I3  Insecure Ecosystem Interfaces           \\ \hline
I4  Lack of Secure Update Mechanism         \\ \hline
I5  Use of Insecure or Outdated Components  \\ \hline
I6  Insufficient Privacy Protection         \\ \hline
I7  Insecure Data Transfer and Storage      \\ \hline
I8  Lack of Device Management               \\ \hline
I9  Insecure Default Settings               \\ \hline
I10 Lack of Physical Hardening             \\ \hline
\end{tabular}
\end{table}
  
  The scope of our problem is detecting the potential data leaks in IoT applications in the form of tainted data flows from tainted sources, which are variables or parts of the code containing sensitive information,  to sinks, which are functions that can leak the information. This happens when certain variables are communicated or pushed through a channel which can be compromised. This can be caused either by carelessness by the programmers or intentionally by an attacker. Programs can falsely provide descriptions and ask for permissions to do certain functionalities which might violate users' privacy without them knowing. This can be done either maliciously or due to bad programming practices. We aim in this research to deal with this issue and provide an approach to check the source code for cases where such leakage \hlc[white]{could potentially} happen. \hlc[white]{Our approach  provides a quicker core analyzer as well as adding sensitivities as modules which can help detect the parts in the code that contains tainted flow and potential leakage. This could help the developer or the reviewer to scrutinize them further and put safety measures in place if needed.}
  
  This paper is an extension to our previous short paper \cite{b0}, where we briefly presented an efficient and scalable static analysis approach and tool, Taint-Things, to identify information leakage in smart things apps. The approach provides security auditing reporting via computing and presenting tainted flow security slice directly from the code using an inductive transformation paradigm \cite{txl}.  
 
  And the main contribution of this extended paper is:
    \begin{itemize}
  \item Extends Taint-Things with flow-, path-, and context sensitivity analysis and that to improve the tool's Precision and Recall.
  \item An experiment that evaluates Taint-Things Precision and Recall after adding flow-, path-, and context sensitivity analysis.
  \item We demonstrate that static analyses on IoT can be pushed to be quicker and cheaper by doing the analysis directly on the source code, with the right tools and without needing a lot of intermediate representation. This gives us the ability to explore ways to further increase in precision with less cost.
  \end{itemize}

Our approach and tool can be used to give more transparency to the user in terms of applications' functionality by reporting tainted flows which are potential areas in the code that contain and may leak sensitive information, as well as providing developers and reviewers with a beneficial tool which can automate the process of detecting tainted data flows. Our approach reports potential vulnerable tainted flow in a form of a concise security slice, that is the relevant parts of the code containing the flow. Our approach and tool could provide security auditors with an effective and precise tool to pinpoint security issues in SmartThings apps under test.  

In the following section, Section \ref{ch:Background}, we provide a background and an overview of some the concepts related to our research, which includes static analysis, sources and sinks, and sensitivities. In Section  \ref{ch:Survey}, we provide a literature survey of publications related IoT security as well as some relevant research done on static analysis in Android apps. In Section \ref{ch:Methodology} we demonstrate our approach and its implementation. The following Section \ref{ch:Precision} we explore ways for adding more precision by handling flow, path and context sensitivities, respectively and evaluating them and on Section \ref{ch:Evaluation} we present an evaluation for our tool and the approaches that add precision. 

\section{Background}\label{ch:Background}
  The scope of this paper is the issue of privacy leakage in IoT apps. We present a method of detecting potentially leaky IoT apps, illustrated by SmartThings apps, by analysing their source code, and trying to see ways of improving the analyses. To start off, we provide some of the important concepts that are used within this study.
  \begin{lstlisting}[style=Groovy, float, floatplacement=t!, caption={SmartThing App structure}, label=list:groovyExample]

    definition(
        name: "Sample App",
        //...
    )
    
    preferences {
        section("User input example:") {
		    input "phone", "number", title: "User inputted number"
        }
    }
	
	//..
    def initialize() {
        subscribe(phone, "number", eventHandler)
    }
    
    def eventHandler(evt) {
        sendSms "Your value is: $phone"
    }
    
    def closureExample(){
        def List numberList = [1, 2, 3]
        numberList.each{print it}
    }
    
    def foo() {...}
    def bar() {...}
    
    def callByReflection(){
        def methodName = "foo"
        $methodName
    }
   \end{lstlisting}

  \subsection{The Platform}\label{sec:Platform}
  IoT has many frameworks, platforms and vendors, and while they might differ in the way they handle permissions and granularity, they share key concepts. Such platforccms include: Samsung's SmartThings \cite{smartthings}, Apple's HomeKit \cite{apple}, openHAB (open source)\cite{openhab}, Vera Control's Vera3 \cite{vera3}, Google's Weave/Brillo \cite{weave}, and Open Connectivity Foundation's AllJoyn \cite{alljoyn}.
  
  \hlc[white]{Celik et al} \cite{b7} \hlc[white]{point to five IoT specific challenges when it comes to the security of the platforms. Namely, the issues of physical channels, simulation and modeling, test generation, multi-app analysis, and  interaction between devices and platform services.
  
  Physical channels is an issue because IoT devices control physical devices, which puts additional security risk through the physical processes. This can be through side-channel leaks, health related risks, and risk from indirect interactions.Side-channel leaks can happen for example when an adversary use the changes in an IoT device to infer whether someone is at home or not, Health related risk is a serious issue with IoT devices, since they control things like temperature or sound. IoT could also be indirect controlled, for example, by altering the environment around it, one can control certain devices.
  
  Simulating IoT platforms is also challenging, since IoT devices often communicate with each other and act as a complex system. This, and the fact that IoT devices interact physically with their environment makes modeling and simulation challenging. This can affect how well one can study the security aspect of the whole system.
  
  Additionally, since IoT is a new field, systematic and automatic test generation is still an issue that is not explored well. This is especially important for dynamic analysis.
  
  IoT apps can also interact with each other; their events could be tied to each other or they can interact with the same device at the same time. This proposes a challenge when analyzing the apps and makes it necessary to consider the behavior of multiple apps at once.
  
  Furthermore, IoT devices can interact with other platforms, such as network APIS or authentication services. This adds to the complexity of the analyses where the platform service and its interactions should be considered in the analyses.
}

  We chose to start our study with a focus on Samsung's SmartThings because it is one of the more mature platforms with a good user base and it shares the important principles with the other platforms. SmartThings has three components: the hub, the apps and the cloud back-end. \hlc[highlight]{For our research, we are concerned with the apps and potential vulnerabilities in their programming. The scope of our research is thus specific to challenges relevant to the app side rather than the whole system.} 
  
  \hlc[white]{SmartThings apps have security measures such as privilege separation, secure storage and apps are written in a sandbox environment, where features are limited for more security. Nonetheless,  the platform raises some security concerns such as WebServices where HTTP endpoints get exposed and the use of call by reflection. This can propose a vulnerability if combined with over-privilege; an attacker can use this to execute command injection  attacks. There is also little restriction on the communication abilities through the internet or through SMS, which can be used to leak sensitive information

  One of the challenges when studying the system is the fact that it is closed-source, uses a proprietary environment for the execution and the system does not have publicly available APIs to obtain binaries. This makes dynamic analysis hard.}
  
  The apps in SmartThings are written in Groovy, a language like Java, and developed in a sandbox which limits it to specific functionalities relevant to IoT development. The usual structure of a SmartThings app is comprised of the following sections: definitions, preferences and the events/actions sections.
  
  The definition section is where the application's name, description, category and other information are described. The preferences section is where permissions are defined for different devices as well as user inputs. Event/action sections define the methods which will perform actions required whenever an event is triggered.
  
  SmartThings apps are event-driven by design, like Android applications, they do not have main functions. Instead, they have events that triggers the method calls. Some of the language-specific unique features include closures and call by reflection. Closures can be used to loop and perform action on a defined list of elements and call by reflections allows for the calling of methods by using their name in strings. Listing \ref{list:groovyExample} shows an example of the structure of a SmartThings app as well as examples of closure and call by reflection usage.

\subsection{Static Analysis}\label{sec:StaticAnalysis}
  Program analysis can be used to solve different problems, whether it is checking for the correctness of a program, finding ways to optimize it or improving its security. Under the umbrella of security, program analysis can be used with IoT apps to detect sensitive data leakages. It can be done either static, without executing the source code, or dynamic, during run time. Static analysis requires access to the source code but has the advantage of providing more coverage and enables us to examine the structure of the code, while dynamic analyses is limited by the scope of the code being executed. 
  
  When it comes to analysis tools, we care about certain attributes such as soundness and completeness. Where soundness deals with the correctness of the reports and completeness deals with covering all what's there to report. Measures used to evaluate correctness and completeness are precision and recall. Precision gives an idea of how many false positives are being reported from the total positives and Recall gives an idea about completeness by calculating how many true positives are reported out of all the positive cases. Making a perfect static analysis tool can be an impossible task, so we depending on the task, we can try to achieve certain features with trade-off from others.
    
  We use a static analysis approach to tackle the problem of data leakage in IoT programs, where we perform the analysis directly on the code without executing it. The goal of the analysis here is to detect the flow from sources of potential sensitive data to sinks which are potential data leakage points in the code. If a flow contains sensitive information, we consider it tainted and we report it.
 
  Different factors might contribute into the analysis precision. Precision being the value of avoiding false positives. In the following subsections, we present an overview of patterns that exist in IoT programs which can affect precision and how an analysis could takes them in consideration, making it a sensitive one:
  
    \subsubsection{Flow Sensitivity}
    A flow sensitive analysis takes in consideration the statement execution order in the program as well as content change in variables. Listing \ref{list:flowSensEx} shows a sample code where the original variable \textit{message} contains sensitive data and is passed to the sink \textit{sendPush} but after it is changed, it is sent to \textit{sendSms}. A flow sensitive analysis takes that into consideration and marks \textit{sendSms} and  \textit{sendPush}'s data flows differently with  \textit{sendSms} being not tainted, since it accurately detects no sensitive data being sent through it, while an insensitive approach will confuse the two and mark  \textit{sendSms} as a tainted sink, since \textit{message} contained sensitive data at one point.
    \begin{lstlisting}[style=Groovy, caption={Flow Sensitivity Example}, label=list:flowSensEx]
    // $sensitiveData defined
    def message = "This contains $sensitiveData";
    sendPush(message);
    message = "no sensitive data"; 
    sendSms(message);
    \end{lstlisting}
    To achieve flow sensitivity we can deal with each changed variable as a new variable all together; each variable would only be assigned a value once.
    
    \subsubsection{Path Sensitivity}\label{sec:PathSensitivity}
    A path sensitive analysis takes the execution path in consideration. This is exemplified in how it deals with conditional statements; a path sensitive analysis would treat each conditional block as a separate path. Listing \ref{list:pathSensEx} illustrates this; a path sensitive analysis would only detect \textit{sendSms} on line 5 as a tainted sink, while an insensitive approach will mark both \textit{sendSms} and \textit{sendPush}.
      \begin{lstlisting}[style=Groovy, caption={Path Sensitivity Example}, label=list:pathSensEx]
    //$sensitiveData defined
    def message = "no sensitive data";
    if (condition) {
        message = "This contains $sensitiveData";
        sendSms(message)
    } else {
        sendPush(message)
    }

    if ($sensitiveData) {
        def newMessage = "no sensitive data";
        sendSms(newMessage)
    }
    \end{lstlisting}
 Another issue related to path sensitivity is implicit flows, which are flows that occur implicitly when a conditional statement depends on sensitive data. In Listing \ref{list:pathSensEx} \textit{sendSms} on line 12 doesn't send any sensitive data, but it is in a conditional path that depends on testing sensitive data. In a path sensitive approach that takes implicit flows in consideration, the whole conditional block would be considered tainted and \textit{sendSms} on line 12 would be declared a tainted sink, while a path insensitive approach or one that doesn't take implicit flows in consideration won't mark it as potentially containing or leaking sensitive information.
    \subsubsection{Field Sensitivity}\label{sec:FieldSensitivity}
    A field sensitive approach differentiates between the fields in an object as apposed to treating them as if they were the same. This is mainly relevant when dealing with global variables in SmartApps; global variables are stored as fields in an external object. Listing \ref{list:fieldSensEx} shows an example of two global state variables. In a field sensitive analysis, each one would be modeled as its own, while an insensitive one will treat them as if they were the same variable throughout the program.
    
    \begin{lstlisting}[style=Groovy, float, caption={Field Sensitivity Example }, label=list:fieldSensEx]
    state.firstCounter = x
    state.secondCounter = y
    \end{lstlisting}

    \subsubsection{Context Sensitivity}\label{sec:ContextSensitivity}
    Context sensitivity mainly deals with function calls and callbacks within the program. A context sensitive analysis identifies each call as its own and can track back to the context of the call.
    
    \begin{lstlisting}[style=Groovy, float, linewidth= 0.9\textwidth, caption={Context Sensitivity Example },label=list:contextSensEx ]
    //$sensitiveData defined
    def takeAction() {
        def message = "This contains $sensitiveData";
        def firstCall = returnMessage(message);
        def secondCall = returnMessage("no sensitive data");
        sendSms(secondCall);
    }

    private returnMessage(message) {
        return message;
    }
    \end{lstlisting}
Listing \ref{list:contextSensEx} shows a case where a method gets called twice. In a context insensitive analysis, both calls on lines 4 and 5 might be conflated, so the tainted return in \textit{firstCall} will also be considered in \textit{secondCall}, marking the flow tainted. A context sensitive approach on the other hand, doesn't confuse method calls and distinguishes each call site, so it won't mark the flow to \textit{sendSms} tainted. 
  \subsection{TXL}
  TXL \cite{txl2} is a programming language that can be used for program  transformation. It is a Functional/Rule-based hybrid language. A TXL program has two main parts, a grammar and transformation rules. The grammar part defines the syntax of the inputted  program and allows for the TXL to recognize and parse its structure, while the transformation rules modifies it by replacement and alteration. Listing \ref{list:TXLGrammar} is an example of a grammar for program made of a simple arithmetic statement; it specifies arithmetic expressions' priorities, their structures and their syntax.  Listing \ref{list:TXLTransformation} is an example of a transformation rule using the provided grammar to replace addition expressions with the result of the operation. The rule tries to find any expression that matches the provided pattern, and replaces the two numbers it matches with their sum. The \textit{number [+ number]} is the built-in TXL expression for addition.
  
  \begin{lstlisting}[ caption={TXL Grammar}, float, label=list:TXLGrammar]
    define program
        [expression]
    end define
    
    define expression
          [term]
        | [expression] + [term]
        | [expression] - [term]
    end define
    
    define term
          [primary]
        | [term] * [primary]
        | [term] / [primary]
    end define
    
    define primary
          [number]
        | [term]
        | ([expression])
    end define
   \end{lstlisting}
   
  \begin{lstlisting}[caption={TXL Transformation}, float, floatplacement=H, label=list:TXLTransformation]
    include "grammar.grm"
    
    rule main
        replace [expression]
          N1 [number] + N2 [number]
        by
          N1 [+N2]
    end rule      
  \end{lstlisting}
  
\subsection{Testing Plan}
  To test our we work we conduct a comparative analysis with the state-of-the-art available tool on a dataset of the available SmartThings apps which are either provided officially or by third-parties by the community. 
  We used a computer with similar specs to what the state-of-the-art the tool used to evaluate the speed and we look into the real time and CPU time taken to analyse the full dataset. We also manually check the reported flows by both tools by looking into the code and confirming that the results are sound.
  
  For the added sensitivities, we also use a mutation dataset, which is based on the original dataset but adds patterns which contain patterns that require flow, path and context sensitive analyses to avoid false positives. We calculate the precision and recall for the results for our tool when adding the sensitive analysis and compare it to the original without it.
\section{Taint Things}\label{ch:Methodology}
  To implement our static analysis approach we use TXL \cite{txl}. The components of a TXL program are a grammar and a set of transformation rules. The grammar specifies the structure of the input while the rules specify the patterns that TXL will detect and replace to produce the output.

  The challenge for the identification of a tainted flow is to trace dependencies backwards in the program to mark only those statements that can influence the marked tainted sinks.

  Static taint analysis techniques, such as SAINT\cite{b5}, build a dependency graph for the program and then use graph algorithms to reduce it to the tainted flow, which is mapped back to source statements afterward. \hlc[highlight]{The idea of using dependency graphs goes back to Thomas Reps. et al.} \cite{10.1145/989393.989419}. However, in our approach we compute dependency chains directly, using the inductive transformation paradigm.  
  
\begin{figure}[t!]
    \centering
    \includegraphics[width=0.9\textwidth]{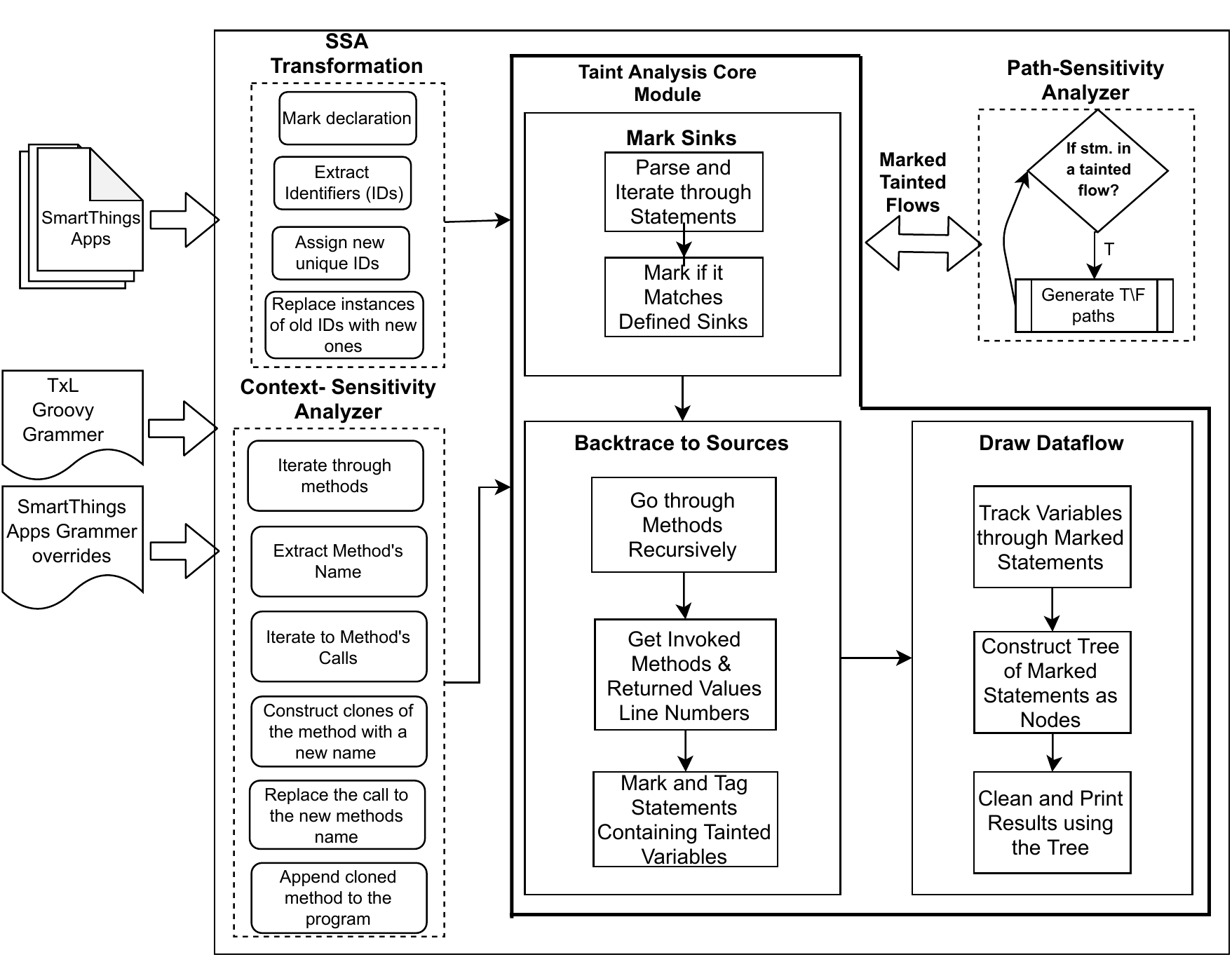}
    \caption{\hlc[highlight]{Taint-Things Approach Architectural Diagram}}
    \label{fig:Approach2}
    \end{figure}

  The approach uses a related TXL paradigm called cascaded markup. \hlc[highlight]{As Figure}\ref{fig:Approach2}-\hlc[highlight]{\textit{"Taint Analysis Core Module"} demonstrates}, the approach starts with marking sink statements, analyzing them recursively, finding and marking statements which directly influence them and then those that influence those statements, and so on until a fixed point is reached. This fixed point occurs when a potential tainted source is identified or when no more propagation can be done. As stated earlier, an example of potential tainted sources are user input identifiers that have been defined as part of the preference block in the SmartThings app. For the cascaded markup, we consider an assignment to a variable to be part of a tainted flow if any subsequent use of that variable exists in a sink. 	

  The other markup propagation rules are simply special cases of this basic rule that propagate markup backwards into loops and if statements, around loops and out to containing statements when an inner statement is marked, \hlc[highlight]{Figure} \ref{fig:Approach2}-\hlc[highlight]{\textit{Path Sensitivity Analysis}}. The whole set of markup propagation rules is controlled by the usual fixed-point paradigm that detects when a tainted source is hit or when no more propagation can be done. The analysis as well takes care of taint propagation via methods calls and returns, \hlc[highlight]{Figure}\ref{fig:Approach2}-\hlc[highlight]{\textit{context sensitivity Analysis}}.
 
    \subsection {Taint Analysis-Core Module:}    
  \subsubsection{Building a TXL Grammar}
  The TXL grammar is considered as the most crucial component. It specifies the way TXL parses the program, like context free grammar (CFG), it analyzes the program into its components, which are the non-terminal statements, and further specifies the components of each statement down to terminal components, such as operators and operand.
  
  Since our first step is to support SmartThings applications, which use Groovy, we must start with a grammar for it. Groovy grammar for TXL is not readily available, so we had to develop it ourselves.  This could be done in two ways; since Groovy is similar in many ways to Java, we could override the available TXL Java grammar\cite{JavaGrammar}, tailoring it to Groovy, or we can start from scratch, writing a Groovy specific grammar.
  
  We tried the first approach. Starting with the Java grammar and trying to parse Groovy applications, patching it up whenever it fails. We found out that as programs got bigger and used more Groovy specific features, the grammar was unable to parse them, resulting into errors, and was getting harder to maintain and patch. We developed and adapted 1400 lines of code and  were only able to get 80\% of the dataset to parse. The lack of semicolons as command separators and string interpolation were the biggest issues. This is because most statements in Java must end with semicolons while in Groovy, semicolons are optional and string interpolation has the challenge of dealing with call by reflection where you can call a method from within a string.
  
    \begin{lstlisting}[style=Groovy, caption={Example of Problematic Statements}]
    int foo = 3
    String b = "Foo is $foo"
    \end{lstlisting}
  
  For the second approach, crafting a TXL grammar, we needed to start with an extended Backus-Naur form (EBNF) for Groovy. However, this isn't provided in Groovy's official documentation. Instead, the official repository offers parser and lex files generated from ANTLR \cite{GroovyAntlr} which can still give us an idea of the structure of the language. We used these as a starting point in our grammar inference process. With this approach, we developed 600 lines of code, and achieved 100\% success rate in parsing our dataset.
  
  \subsubsection{Identifying Sinks}
  Taint sources are defined as variables or information which get passed in the application and potentially contain sensitive information. On the other hand, sinks are defined as functions which pass the information and potentially leak it. In our approach, we identify the grammatical forms for all potential sources and sinks for SmartThings IoT app described by Celik et al.\cite{Celik}. Their list includes the following sources: device states, device information, location, user inputs, and State variables. Listing \ref{list:userInputtedVar} shows an example of a defined user input in a SmartThings app.

    \begin{lstlisting}[style=Groovy, caption={Example of a user inputted variable in a SmartThings app }, label=list:userInputtedVar]
        input "myLock", "lock", multiple:true, necessery:true
    \end{lstlisting}
    
     \begin{lstlisting}[caption={TXL Grammar Definition for Sinks}, label=list:sinksGrm]
    define sink_name
          'httpDelete | 'httpGet | 'httpHead | 'httpPost 
        | 'httpPostJson | 'httpPut| 'sendSms | 'sendSmsMessage 
        | 'sendNotificationEvent | 'sendNotification
        | 'sendNotificationToContacts|'sendPush| 'sendPushMessage 
    end define
    \end{lstlisting}
    
    Then we add those patterns to the grammar description of our analysis tool. Listing \ref{list:sinksGrm} provides an example of how we grammatically define sinks functions used in SmartThings apps and identified by our tool. This approach provides flexibility by allowing us to add or remove potential sources or sinks; to do that, we can simply modify the grammar description for sources and sinks patterns. This change will not impact our analysis, since our rule based pattern matching engine will match by the pattern category rather than the individual patterns elements. 
    
     The first step of our approach is identifying the the sinks in a program. Since sinks are limited and usually less than the sources, it is generally easier to start with them and do backward tracing from there. In this step our tool \hlc[highlight]{parses the program and iterates through its statements. If a statement contains one of the defined sink functions, it labels the sink function. It also tags every variable and method declaration with their respective line numbers}. Listing \ref{list:sinksMark} shows an example based on a SmartThings apps and shows how it gets processed then at this step the sink, \emph{sendPush(message)} gets marked with the label \emph{$<sink>..</>$}. Not that when the source code is inputted in TXL, identifiers are labeled with their line numbers.
    
    \begin{lstlisting}[style=Groovy, firstnumber=32, float, floatplacement=H, caption={Sinks Marking Output},label=list:sinksMark]
    def <@\textcolor{line-numbers}{32}@> initialize () {
        <@\textcolor{line-numbers}{33}@> subscribe (<@\textcolor{line-numbers}{33}@> themotion, " motion.active ", <@\textcolor{line-numbers}{33}@> motionDetectedHandler)
    }
    def <@\textcolor{line-numbers}{36}@> motionDetectedHandler(<@\textcolor{line-numbers}{36}@> evt) {
        def <@\textcolor{line-numbers}{37}@> message = " motionDetectedHandler called : 
        $ <@\textcolor{line-numbers}{ 37}@> evt "; 
        <@\textcolor{red}{<sink>}@> <@\textcolor{red}{sendPush}@>(<@\textcolor{line-numbers}{38}@> message) <@\textcolor{red}{</>}@>;
        <@\textcolor{line-numbers}{39}@>
        theswitch.<@\textcolor{line-numbers}{39}@> on ()
    }
  \end{lstlisting}
  \subsubsection{Doing Backward Tracing}
  In the second step of the analysis process, our approach tracks the variable that is being passed to the sink, traces it and marks the lines where it's contained all the way to a source. \hlc[highlight]{This is done by recursively going through the methods, extracting the line numbers of invoked methods and line numbers of returned values and using them to track and mark statements that contain a tainted variable, tagging it with the line number where it gets passed to.}
  
  In the following listing \ref{list:bTracingOutput}, the variable \textit{message} is passed to the sink. And \textit{message} has another variable \textit{evt} which is passed to the function \textit{motionDetectedHandler}. In the program initialization this function is subscribed with the source user input \textit{themotion}. The subscribe statement is tagged with the line number where the function is defined.
  
  \begin{lstlisting}[style=Groovy, float, floatplacement=H, firstnumber=32, caption={Back Tracking Output},label=list:bTracingOutput]
  preferences {
    12 section(" Turn on when motion detected: ") {
      input " themotion ", " capability.motionSensor ", 13 required : true, 13 title : " Where ? "
    } 
    ...
  } 
  def   <@\textcolor{line-numbers}{32}@> initialize () {
    <@\textcolor{tags}{<36>}@>  <@\textcolor{line-numbers}{33}@> subscribe (   <@\textcolor{line-numbers}{33}@> themotion, " motion.active ",   <@\textcolor{line-numbers}{33}@>  <@\textcolor{red}{motionDetectedHandler)}@> <@\textcolor{tags}{</>}@>
  }
  def   <@\textcolor{line-numbers}{36}@> motionDetectedHandler( <@\textcolor{tags}{< >}@> <@\textcolor{line-numbers}{36}@> <@\textcolor{red}{evt}@> <@\textcolor{tags}{</>}@>) {
    def <@\textcolor{line-numbers}{37}@> message = " motionDetectedHandler called : 
    $ <@\textcolor{line-numbers}{37}@> <@\textcolor{tags}{< >}@> <@\textcolor{red}{evt}@> <@\textcolor{tags}{</>}@> "; 
    <@\textcolor{red}{<sink>}@>  sendPush(  <@\textcolor{line-numbers}{38}@>  <@\textcolor{red}{message}@>)   <@\textcolor{red}{</>}@>;
    <@\textcolor{line-numbers}{39}@> theswitch.  <@\textcolor{line-numbers}{39}@> on ()
  }
  \end{lstlisting}
  
  \subsubsection{Identifying Tainted flow}
  In the next step of the analysis, the data flow gets drawn. To do that, the previously marked lines are tagged with the line numbers where their variables gets passed. Those get used in determining the feasibility of the flow \hlc[highlight]{by constructing a tree from the tagged statements as nodes}. For the report, the code is cleaned by removing \hlc[highlight]{unmarked statements. The report is generated using the constructed tree for the line numbers containing tainted dataflow. Finally,} variable and method declaration line numbers, tagging the source and presenting a summary of the data flow. The output from Taint-Things is shown in listing \ref{list:securitySlice} where the numbers on top are the summary representing the line numbers of the flow and the source code gives the details to where it exists. \hlc[highlight]{An example of a real SmartThings app and the analysis steps is provided in the Appendix} \ref{appendix:a}.
  
  \begin{lstlisting}[style=Groovy, float, floatplacement=H, caption={Taint Flow Report},label=list:securitySlice]
  <@\textcolor{red}{33 36 37 38}@>
  
  def initialize() { 
    <<@\textcolor{red}{ 36 source }@> > subscribe(themotion, " motion.active ", motionDetectedHandler) < / >
  }
  
  def motionDetectedHandler( <<@\textcolor{red}{37}@>> evt < / >) {
    < <@\textcolor{red}{38}@> > def message = " motionDetectedHandler called : $ evt "; < / >
    < sink > sendPush(message) < / >
  }
  \end{lstlisting}

\section{Precision Enhancement}\label{ch:Precision}
  While we've seen that the core module of Taint-Thing by itself is satisfactory in terms of detecting potential leakages as the tests on the dataset show, we wanted to see if there are ways to improve it by making it more resistant to potential false positives and more precise. Since it is fast and light in its performance, in this and the following sections, we want to explore how we can improve the precision by adding analyses sensitivities. This can help us minimize false-positives in certain cases. In this section we'll look into the case of flow sensitivity which is relevant in the case of variable reassignment throughout the code. 
  \begin{figure}[b!]
    \centering
    \includegraphics[width=0.5\textwidth]{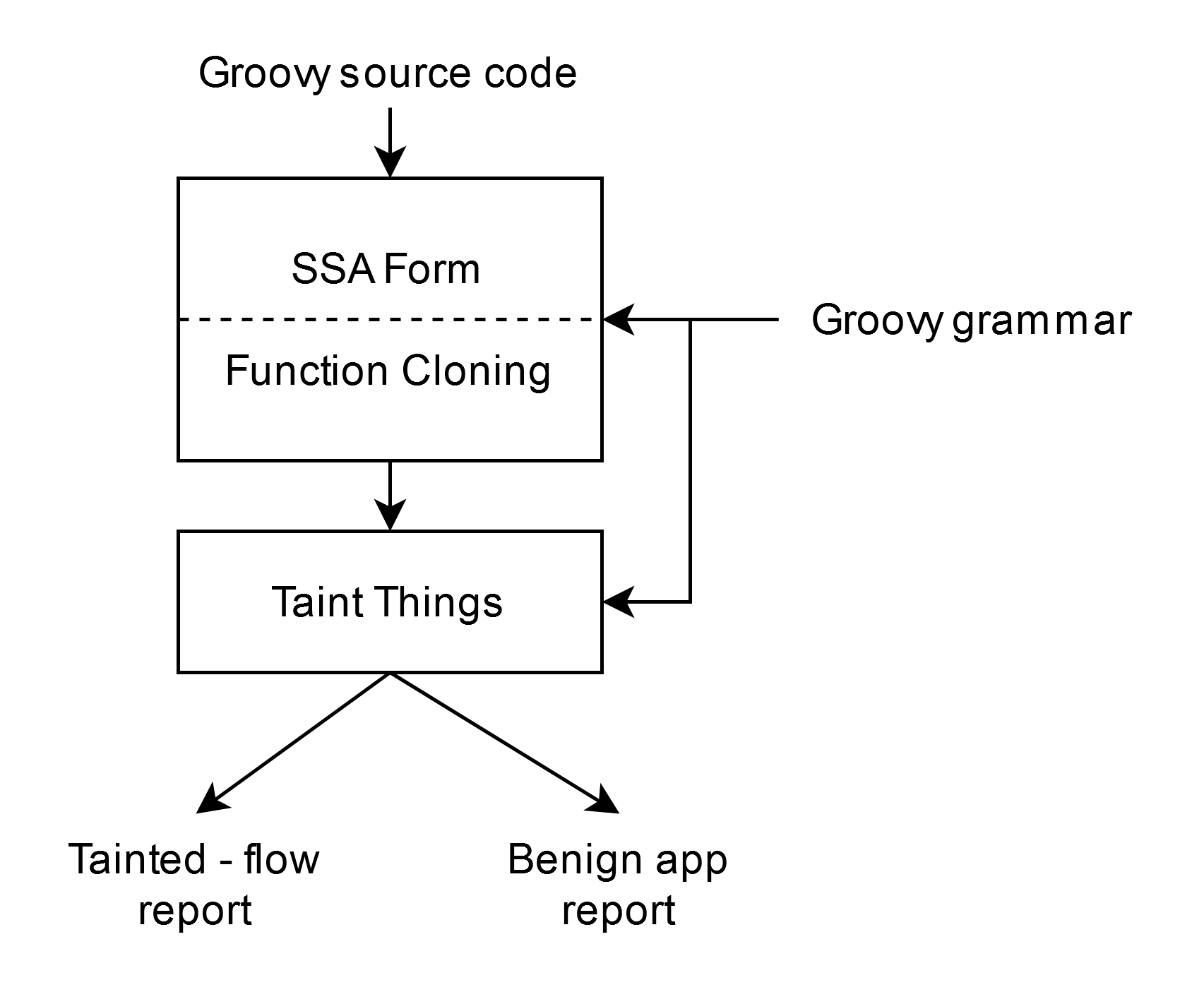}
    \caption{Taint-Things with SSA or Function Cloning}
    \label{fig:ApproachB2}
    \end{figure}
  \subsection {Adding Flow Sensitivity}\label{ch:Flow}
  A flow sensitive analysis takes in consideration the order of statement execution in the program as well as content change in variables. A flow sensitive analysis takes the variable's value changes into consideration while an insensitive approach does not. This can result into false positive reports. Currently, this is the case with both Taint-Things and SAINT, if we test the program in listing \ref{list:flowSensEx2} they both mark the flow as malicious even though it is benign. 
  
  \begin{lstlisting}[style=Groovy, float, floatplacement=H, caption={Flow Sensitivity Example}, label=list:flowSensEx2]
    // $sensitiveData defined
    def message = "This contains $sensitiveData";
    sendPush(message);
    message = "no sensitive data"; 
    sendSms(message);
    \end{lstlisting}
  To avoid conflation in this case and achieve flow sensitivity, we can deal with each variable change as if it were a new variable; each variable would only be assigned a value once. This is the same concept used in static single assignment (SSA) form where each variable in the code is assigned a value once. Listing \ref{list:SSAex1} shows Listing \ref{list:flowSensEx2} in SSA form. We can process the inputted program and transform it into this form, which we later can chain into Taint-Things. This form will provide more precision when dealing with variable reassignments and will solve the issue of false positives resulting from them. 
  
  \begin{lstlisting}[style=Groovy, caption={SSA form Example }, label=list:SSAex1]
  // $sensitiveData defined
  def message1 = "This contains $sensitiveData";
  message2 = "no sensitive data"; 
  sendSms(message2);
  \end{lstlisting}
  
  We have implemented this approach and written a TXL program that does this transformation. Using the groovy grammar, we analyse and detect all the assignment statements in a program as well as the identifiers. \hlc[highlight]{ The program marks all declaration statements in the inputted program. It then gets the variable's identifier from each declaration statement and assigns a unique variable identifier for that declaration statement. Finally, it replaces any instance of the old variable with the newly assigned identifier.} Listing \ref{list:assignemtInFSE} shows the detected assignment statements in Listing \ref{list:flowSensEx2}.
  
  \begin{lstlisting}[style=Groovy, float, floatplacement=H, caption={Assignment Statements }, label=list:assignemtInFSE]
  // $sensitiveData defined
  *def message = "This contains $sensitiveData";*
  *message = "no sensitive data"*
  sendSms(message);
  \end{lstlisting}
\begin{algorithm}
\KwIn{A program's source code}
\KwOut{Program's source code in SSA form }
\BlankLine
\SetKwFunction{FMain}{generateSSA}
\For{each variable $V$ } {	
    $C(V) \leftarrow 0$\\
}
    
\For{each statement $A$ in program $X$ } {	
    \If{$A$ is an assignment}{
        get $V$ from $LHS(A)$\\
        $i \leftarrow C(V)$\\
        $C(V) \leftarrow i + 1$\\
        replace $V$ by new $Vi$ in $LHS(A)$\\
        \For{statement $B$, from $A$ to $end$}{
            \If{$B$ contains a variable that is equal to $V$}{
                $i \leftarrow C(V)$\\
                replace by new $Vi$
            }
        }
    }
}
\caption{Rename variables in a program to adhere to SSA form}
\label{algo:SSA}
\end{algorithm}
     
  Algorithm \ref{algo:SSA} represents this process. Starting from top to bottom. We lookup identifiers that match the one defined in the assignment statement, then we give that identifier set a new unique variable name. Repeating this processes, we go through each assignment statement and its associated identifiers. The result will be a program in SSA form, where every assignment statement gives us a uniquely named variable and avoids conflation in the case of reassignment. This can later be chained with our analysis tool, giving us a more precises analysis. 

  Going through the program's statements, if we encounter an assignment statement, we extract the variable \textit{V} from the left-hand side (LHS). We assign a counter \textit{C(V)} for it and use that to assign a new unique variable name for it. We then go through the rest of statements and if we find a statement containing the original variable, we replace it with the new variable. After that loop is finished, we go to the next assignment statement, and if it contained the same variable name, it will increment its counter, otherwise it'll be assigned a new counter.
  
  \subsection{Adding Path Sensitivity}\label{ch:Path}
  Following the previous section, we look in the case where the flow goes through a conditional statement, where it branches into different paths. We introduce path sensitivity and look in how we can achieve it by taking all possible branches in consideration.
  
\begin{figure}[t!]
      \centering
      \includegraphics[width=0.7\textwidth]{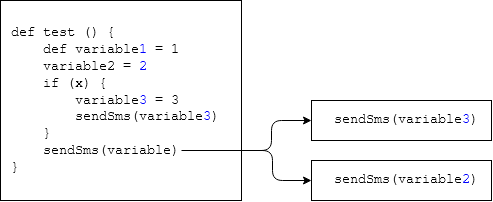}
      \caption{Example of potential paths from an if-statement}
      \label{fig:paths}
  \end{figure}

  A path sensitive analysis takes the execution path in consideration. This is exemplified in how it deals with conditional statements; a path sensitive analysis would treat each conditional block as a separate path. This is important when trying to convert the program into SSA form. Figure \ref{fig:paths} showcases this where the variable sent through the second \textit{sendSms} could either be \textit{variable2} or \textit{variable3} depending on whether the if-statements is executed through the true branch or the false branch.
 
  This is a challenge when using SSA form to deal with flow sensitivity for such cases. The problem is exemplified when dealing with conditional statements, such as if-statements and switch statements. In our thesis we're going to focus on if-statements. These present branching in the program's execution path. A precise analysis takes these branches in consideration and provides all possible cases where a leakage can occur while avoiding false positives.

  In its simplest form, this problem can be represented with a single if statement. This will give us two branches; a branch when true and branch when false. To achieve that and generate these paths from an inputted program, we wrote two TXL transformations. One that removes the if statement, while preserving the new lines. And another that extracts the statements from its body.
  
  At this point, we have the base case solved, but a program could be more complex; it can contain an if-else, multiple if statements, nested if statements or a combination of these. We want to generate all the possible branches from these. To solve this problem, we can recursively generate the true and false paths from the program, removing the if-statements in the process, until we reach the base case.
  
\begin{figure}[t!]
      \centering
      \includegraphics[width=0.5\textwidth]{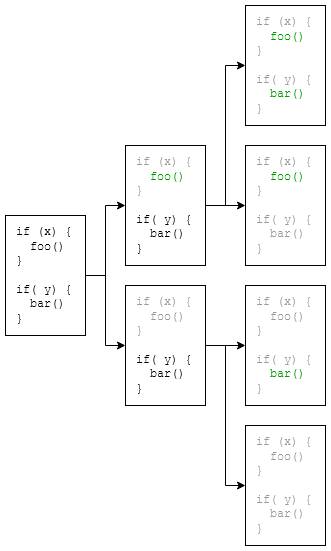}
      \caption{Example of path generation from two if-statements. \hlc[highlight]{Grey lines are deleted while green lines are kept. Top branching represent a True path and bottom branching represent the False path.}}
      \label{fig:2ifpaths}
  \end{figure}

  To do this, we wrote a python script. The script would be responsible for recursively calling the TXL transformation for generating the true and false paths for one if-statement at a time, taking the output and rerunning the transformations, until no if-statement is left. This way we can generate all possible paths from any program with multiple or nested if-statements. After this stage, we would have gotten all the possible paths. Figure \ref{fig:2ifpaths} shows an example of how the program would run on a program with two if statements, generated all the possible four paths.
  
  For if-else statements, for the initial implementation we were converting else statement into if-statements and generating the paths from there using the same method. This approach would end up generating four paths for an if-else statement, with two impossible paths, the path of executing both the if and else clauses and the path where neither is executed. In our later implementation, we changed the way we deal with them so that if we consider the path where the \textit{if} clause is true, the \textit{else} clause is always skipped, and if we consider the path where the \textit{if} clause is false, the \textit{else} clause is always executed. This accurately produces two execution paths for the if-else statements.
  
   \begin{lstlisting}[style=Groovy, linewidth= 0.9\textwidth, caption={Path Sensitivity Example },label=list:pathListing ]
  //sensitiveData defined
  def test () {
    if (x) {
        message = sensitiveData 
    } else {
        message = "benign" 
    } 
    sendSms (message)}
  \end{lstlisting}
  
  As for the final output, each program would give us multiple paths. Each path would be represented by a combination of \textit{T} and \textit{F} symbols. \textit{T} for true path and \textit{F} for false. And the python script would connect these outputs with the SSA transformation and Taint-Things to preform the analysis. To give an example, if we take listing \ref{list:pathListing} as an input, the output produced by our path sensitivity framework is batch report for the two possible paths as showed in listing \ref{list:pathOutput}. We can see both possible paths generated and that one path has a tainted flow reported while the other is benign. 
  
    \begin{lstlisting}[numbers=none, float=ht!, linewidth= 0.9\textwidth, caption={Path Sensitivity Output },label=list:pathOutput ]
  Running Test.groovy
    Path: T

        def test () {
            < 8 source > message1 = sensitiveData < / >
            < sink > sendSms (message1) < / >
        }
    
        <@\textcolor{blue}{4 8}@>
    
    Path: F
    
        def test () {
            < 8 > message2 = " benign " < / >
            < sink > sendSms (message1) < / >
        }
  \end{lstlisting}
  
  \hlc[highlight]{The full implementation can be described as follows. The goal of this module is generating the possible execution paths in the program by generating the true and false branches from if-statements. 
  This is done by running the analyzer to mark statements that are within potentially tainted flows. To check if an if-statement exists that contain a tainted statement,  the module parses the inputted program statements and matches if-statements. It then deconstructs them to see if they contain a marked statement. If such if-statement exists, a TXL program that generates the true path and a TXL program that generates the false path are run. A false path is generated by parsing the program statements, matching if-statements, replacing its content with newlines and deconstructing else-statements and replacing it with the contained statements. A true path is generated by deconstructing the if statements and replacing it with the contained statements and replacing the else-statement with new lines. The program then recursively repeats the process from the output of the path generation until no if-statements are left. Each output is then cleaned from statements markings and is passed to be analyzed.} 
     
 \begin{figure}[t!]
    \centering
    \includegraphics[width=0.5\textwidth]{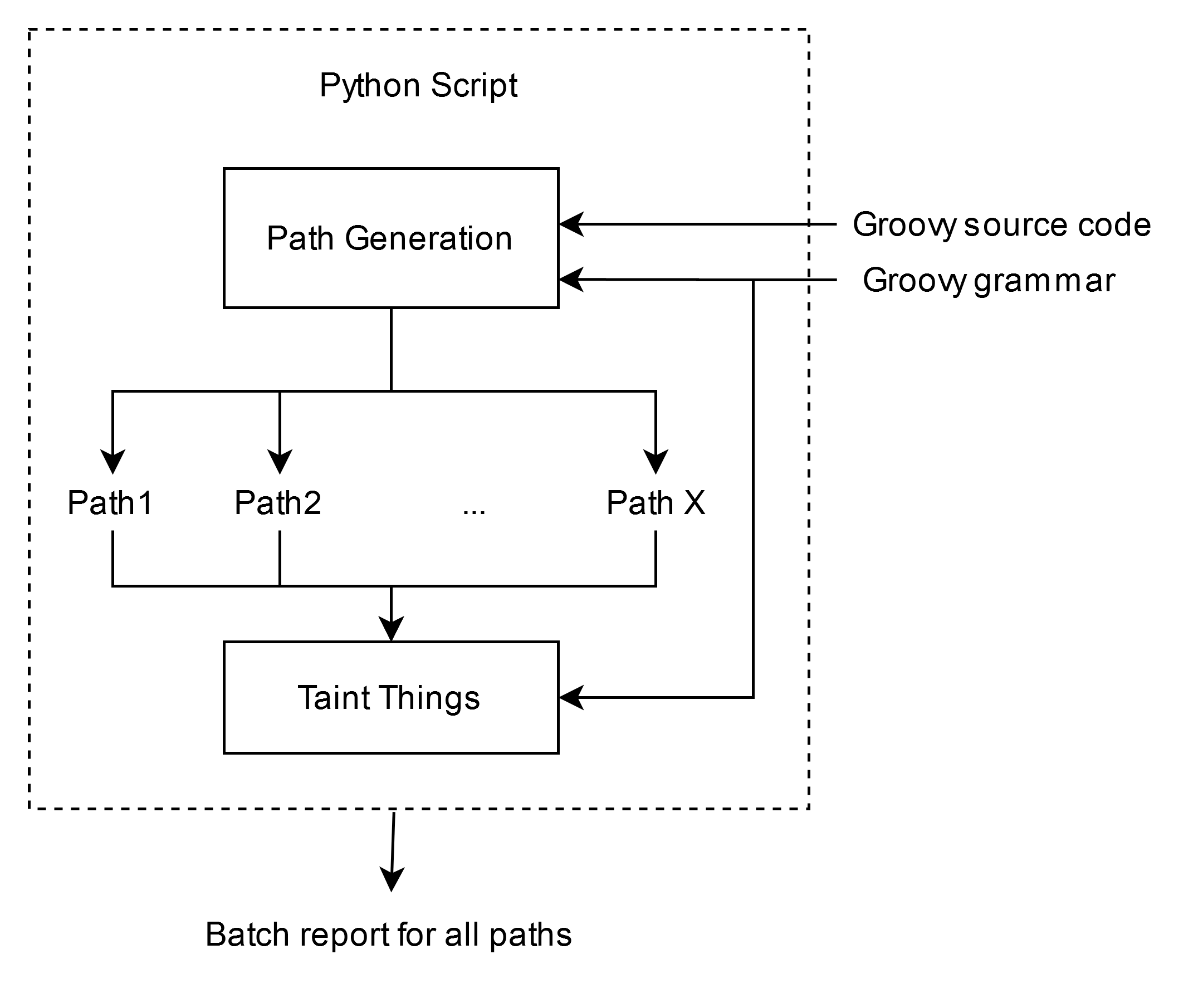}
    \caption{Taint-Things with path sensitivity}
    \label{fig:ApproachB3}
    \end{figure}
  \begin{algorithm}
\KwIn{A program's source code}
\KwOut{Multiple programs representing the possible execution paths}
\BlankLine
\SetKwFunction{FMain}{handleIf}
\SetKwProg{Pn}{Function}{:}{end}
\Pn{\FMain{X}}{
    \If{$X$ contains an if-statement}{
        $truePath \leftarrow generateTruePath(X)$\\
        $falsePath \leftarrow generateFalsePath(X)$\\
        
        $handleIf(truePath)$\\
        $handleIf(falsePath)$\\
    }
}
\caption{Path Generation}
\label{algo:extSourcesMethodInvocs}
\end{algorithm}    
  
\begin{figure}[htp]
    \centering
    \includegraphics[width=1\textwidth]{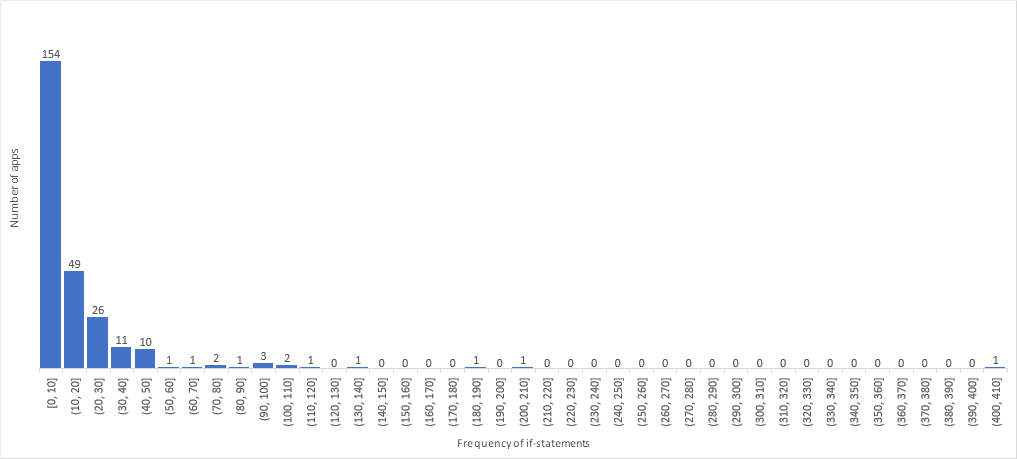}
    \caption{If-statement frequency in the dataset apps}
    \label{fig:ifFrequency1}
  \end{figure}

\begin{figure}[htp]
    \centering
    \includegraphics[width=0.75\textwidth]{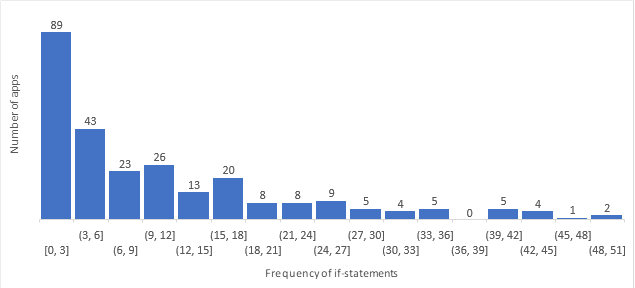}
    \caption{If-statement frequency without outlier values}
    \label{fig:ifFrequency2}
  \end{figure}

  The first thing to note is that this process can be costly in terms of performance. As we have seen, with each if-statement, the program would have double the branching. This grows exponentially with each if-statement added and eventually means that the analysis will be done 2\textsuperscript{n} the times of how many if-statements there are in the program. To illustrate the frequency of if-statements in each of the dataset's apps, in each app we counted how many if-statements exist using a TXL program, then organized the results in a histogram showing how many apps have a frequency of if-statements.  Figure \ref{fig:ifFrequency1} shows the histogram frequency of if-statements in the dataset's apps. We noted 15 apps having over 50 if-statements and can be considered outliers. We excluded these 15 apps in figure \ref{fig:ifFrequency2} to show a better detail of the frequency distribution. In our tests we've seen that a program with more than 12 if-statements becomes too large to add path sensitivity in a practical way. And since the growth is exponential, with each if-statement making at least two possible branches, we want to optimize the process and try to deal with as little if-statments as possible.
  
\begin{figure}[htp]
    \centering
    \includegraphics[width=1\textwidth]{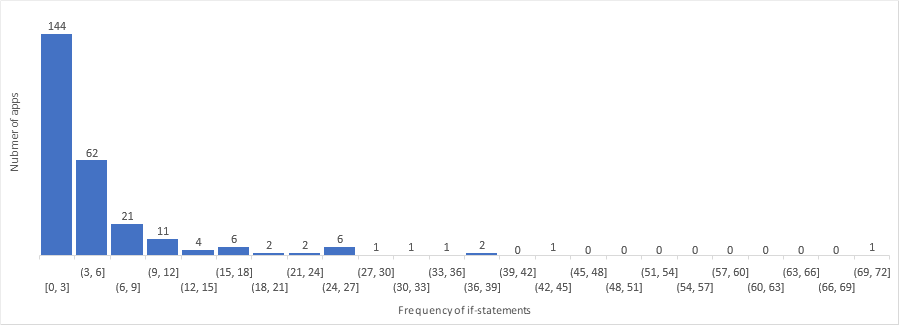}
    \caption{If-statement frequency with method-centered optimization}
    \label{fig:ifFrequency3}
  \end{figure}

  One way of optimizing this is doing the transformations on all the methods at once. So, a program with two methods, each with one if-statement, would only produce two paths, instead of four. So the amount of paths generated would be the exponential of the maximum number of if-statements in one method rather than the total of if-statements in the source code. Figure \ref{fig:ifFrequency3} shows a histogram of the frequency of the maximum if-statement in one method in each app in the dataset. We can see this effectively lowering the amount of possible branches, with less apps having over 12 if-statements that we have to deal with, but this can come on the expense of context-sensitivity and sometimes ignores certain possible paths that affects a flow through multiple methods; this can happen if two methods, dependant on each other, containing if-statements that affect the same data flow.
  
\begin{figure}[htp]
    \centering
    \includegraphics[width=1\textwidth]{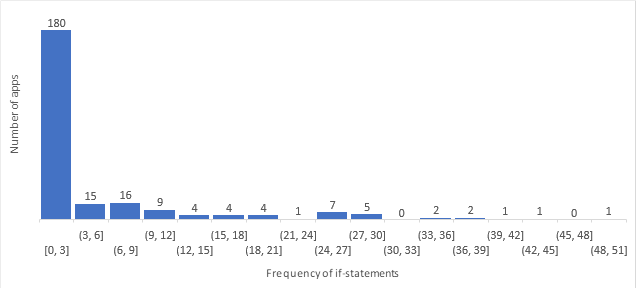}
    \caption{If-statement frequency with method-centered optimization}
    \label{fig:ifFrequency4}
  \end{figure}

  Another thing to note is that not all paths generated necessarily affect the dataflow, so some programs will have multiple paths with the same dataflows reported on different paths. One way to overcome this and provide better optimization is to only consider if-statements that affect the flow. To do that, we can perform a path-insensitive analyses to mark the sinks and backtrace, marking statements in the data flow, then performing the path sensitivity analyses where we generate paths only from if-statements containing the marked statements, and finally performing the analyses again on the paths to generate results. Adding two analysis steps is considerably cheaper than the exponential cost of performing the analyses on all if-statements that don't effect the flow. Figure \ref{fig:ifFrequency4} shows the frequency of if-statements that actually affect the flow in the apps dataset's app. It shows that it can significantly reduce the required amount of if-statements to process in most apps. 
  
 \subsection{Adding Context-sensitive Analysis}\label{ch:Context}
 
  The previous two subsections, we looked into how the flow can affected through variable reassignment and how conditionals can affect that. But sometimes variable reassignment is done through a method calls. Handling method calls and their return values with precision can also pose a challenge. In this section we look into this case where the flow goes through method calls and how to handle the context of each call. 
  
  A context sensitive analysis takes the context of each method call in consideration while a context insensitive approach might confuse them and mark every call to a method as tainted if at one point there were one tainted call. If we take listing \ref{list:contextSensEx} where a method gets called twice, once with a tainted flow passing through it and another with a benign flow. In a context insensitive analysis, both calls on lines 4 and 5 might be conflated; the method is marked tainted after \textit{firstCall}, so \textit{secondCall} will also be considered tainted and the flow tainted. A context sensitive approach on the other hand, doesn't confuse method calls and distinguishes each call site, so it won't mark the flow to \textit{sendSms} tainted. 
    
  \begin{lstlisting}[style=Groovy, float, floatplacement=H, linewidth= 0.9\textwidth, caption={Context Sensitivity Example },label=list:contextSensAddedEx ]
  //$sensitiveData defined
  def takeAction() {
      def message = "This contains $sensitiveData";
      def firstCall = returnMessage(message);
      def secondCall = returnMessage("no sensitive data");
      sendSms(secondCall);
  }
  private returnMessage(message) {
     return message;
  }
  \end{lstlisting}
  
\begin{lstlisting}[style=Groovy, float, linewidth= 0.9\textwidth, caption={Function Cloning Example },label=list:contextSensAdded ]
  //$sensitiveData defined
  def takeAction() {
      def message = "This contains $sensitiveData";
      def firstCall = returnMessage1(message);
      def secondCall = returnMessage2("no sensitive data");
      sendSms(secondCall);
  }
  
  private returnMessage(message) {
      return message;
  }
  private returnMessage1(message) {
      return message;
  }
  private returnMessage2(message) {
      return message;
  }
  \end{lstlisting}
\begin{algorithm}
\KwIn{A program's source code}
\KwOut{Program's source code with function cloning }
\BlankLine
\SetKwFunction{FMain}{generateSSA}
\For{each method $M$ } {
    $C(M) \leftarrow 0$\\
    get method name $MN$ from $M$\\
}
    
\For{each statement $MC$ in program $X$ } {	
    \If{$MC$ is a method call and $MC$ equals $MN$}{
        $i \leftarrow C(M)$\\
        $C(M) \leftarrow i + 1$\\
        replace $MC$ with new $MCi$\\
        copy $M$ body and replace $MN$ with $MCi$
    }
}
\caption{Rename method calls in a program and make clones of the original method with the new name}
\label{algo:Context}
\end{algorithm}

One way to add the sensitivity is through method cloning. Where, similar to the SSA approach, we assign each method call a unique call name and clone the function with it. This assures that each call sight is treated uniquely and no conflation happens. This can be costly if there is a lot of nesting in the methods and won't work properly in the case of recursive functions. Another disadvantage of this approach is that it affects the line numbering of the data flow; depending on where the cloned functions are placed in the source code, the data flow will go through them rather than the original methods. If we apply that to the previous listing, the result will be like listing \ref{list:contextSensAdded}.

  We wrote a TXL program that can be used to do method cloning \hlc[highlight]{in the program, by identifying any method call, assigning a unique method name to it and cloning the method definition using the new name. This is done by parsing the inputted program and iterating through defined methods. Extracting the method's name, parsing the program, and for any expression with a method call that matches the method name, constructing a new unique method name. It then constructs a clone of the defined method but assigns the unique method name to it, and replaces the call to the old method with a call to the new name. Finally, it appends the program with the newly constructed cloned method}


\section{Evaluation}\label{ch:Evaluation}
  \subsection{Evaluating Taint-Things}
  \hlc[highlight]{ In this experiment we try to answer the following two research questions:}
    \begin{itemize}
        \item   \hlc[highlight]{RQ1: How does our static taint analysis approach compares with other available approaches in terms of accuracy and performance?}
        \item   \hlc[highlight]{RQ2: Whether our approach added modules for flow, path and context sensitivity minimize false positives?}
    \end{itemize}
  \hlc[highlight]{To answer \textit{RQ1}, and to measure} Taint-Things's performance in terms of correctly detecting apps containing potential leakage and the speed of the process, we have conducted a comparative analysis experiment with SAINT tool \cite{saint}. We collected SAINT's analysis reports on our dataset and then manually compared their findings with ours. The dataset included 264 applications; 42 official SmartThings Marketplace apps, 144 official apps provided by the community, 59 third-party apps collected from the forums and 19 apps specifically developed by the SAINT team to include common vulnerabilities available online under IoTBench test suite \cite{IoTBench}.

  A thing to note in our comparative study is that some applications would timeout on each of the tools, whether it's due  big size or convoluted function calls. SAINT's web portal having around 42 cases while our tool had 18. For this part of the comparative analysis, we excluded all cases of timeouts.
  
 Table \ref{tab1:resultssum} presents the summary of our comparative analysis with SAINT. The table presents the number of apps that were reported malicious, due to potential leakage of sensitive information in the form of at least one tainted data flow, vs. those that were identified benign. Our results matches SAINT's results in terms of finding which apps have malicious flows except in 5 apps where we report them containing potential tainted flows while SAINT reports them as benign and doesn't report any potential leakage. The criteria for this comparison was looking at what sinks SAINT was reporting and checking if it reported a potential flow being passed to those sinks. 
  \begin{table}[!t]
\centering
\caption{Comparative analysis of SainT and Taint-Things }
\begin{adjustbox}{width=0.6\textwidth,totalheight={\textheight},keepaspectratio}
\begin{tabular}{|l|l|l|l|l|}
\hline
\textbf{App Category} &  \multicolumn{2}{|l|}{\textbf{Taint-Things}} & \multicolumn{2}{|l|}{\textbf{SAINT}}\\
\hline
{}   & Benign   & Leaking    & Benign   & Leaking\\
\textbf{Community (Total: 117)}   &  68 & 49   & 70  & 47\\
\textbf{Forum (Total: 41)}   &  16 & 25   & 17  & 24\\
\textbf{Market (Total: 36)}   &  11  &  25   & 13  & 23\\
\textbf{Benchmark (Total: 19)}   &  1  &  18   & 1  & 18\\
\hline
\multicolumn{4}{l}{\footnotesize * A leaking app is an app with at least one tainted flow is detected}\\
\end{tabular}
\end{adjustbox}
\label{tab1:resultssum}
\vspace{-0.5 cm}
\end{table}

  In addition to comparing our results with SAINT, we verified that our findings were accurate by checking the source code and seeing that the reports matched it. For the cases with mismatched results, we found that the the reason for this is due to lack of field sensitivity in our tool and how we handle state variables. In our tool, we considered any state variable to be a potential source and so mark the flow from it as tainted. This can be addressed on the grammar level and how the program parses state atomic variables. For this case our tool has a more generalized approach when detecting flows from state variables, resulting in the apps that our tool marked and SAINT did not. When looking in the detailed results and comparing the flows reported in each app we've also found that SAINT can report some hard-coded strings as potential leakage sources if they resemble a number, where our app considers hard-coded strings as benign. This can be added by using regex to detect strings that contain phone number patterns or utilize natural language processing to detect strings that can act as sensitive data sources, but at this point, this is beyond the scope of our research.
  
  Table \ref{tabl:DetTaintSaint} shows the reports provided by the two tools, Taint-Things and SaINT when running the benchmark set. Here, we provided the reports as is, but it should be noted that the way the two tools report their results are different, hence the need for manually checking each case. For example, SAINT's report starts with all the detected sinks and all flows, regardless if they were tainted or not, and reports the findings based on the source variables, while Taint-Things only reports tainted flows in the summary and bases the report on the line numbers. When manually checked, we found that the actual tainted flows detected generally match between the two tools with the exceptions mentioned previously, which are state variables and hard coded strings.
  \begin{table}[t!]
\caption{Comparison of the benchmark set between Taint-Things and SAINT}
\begin{adjustbox}{height=0.9 \height, keepaspectratio}
 \label{tabl:DetTaintSaint}
\begin{tabular}{|l|l|l|l|}
\hline
\textbf{} & \textbf{Taint-Things}  & \textbf{SAINT} \\ \hline
\textbf{call\_by\_reflection\_1} & \begin{tabular}[c]{@{}l@{}}Warning followed by \\ 10 tainted flows to 4 sinks \end{tabular} & 5 flows, 4 sinks and   2 findings \\ \hline
\textbf{call\_by\_reflection\_2} & 2 tainted flows to 2 sinks & 3 flows, 2 sinks and 2 findings\\ \hline
\textbf{call\_by\_reflection\_3} & 7 tainted flows to 3 sinks & 7 flows, 2 sinks and 7 findings \\ \hline
\textbf{explicit} & 6 tainted flows to 3 sinks & 5 flows, 3 sinks and 3 findings  \\ \hline
\textbf{global\_variable\_1} & 7 tainted flows to 3 sinks & 5 flows, 3 sinks and 4 findings  \\ \hline
\textbf{implicit\_1} & 3 tainted flows to 2 sinks & 5 flows, 4 sinks and 4 findings  \\ \hline
\textbf{implicit\_2} & 4 tainted flows to 3 sinks & 6 flows, 3 sinks and 4 findings  \\ \hline
\textbf{implicit\_explicit} & 9 tainted flows to 3 sinks & 6 flows, 3 sinks and 8 findings  \\ \hline
\textbf{leaking\_via\_closures} & 5 tainted flows to 3 sinks & 5 flows, 3 sinks and 6 findings  \\ \hline
\textbf{multiple\_devices\_1} & 2 tainted flows to 2 sinks & 7 flows, 3 sinks and 5 findings  \\ \hline
\textbf{multiple\_devices\_2} & 2 tainted flows to 1 sink & 7 flows, 3 sinks and 6 findings  \\ \hline
\textbf{multiple\_devices\_3} & 2 tainted flows to 1 sink & 5 flows, 3 sinks and 5 findings  \\ \hline
\textbf{multiple\_entrypoint\_1} & 11 tainted flows to 3 sinks & 7 flows, 3 sinks and 9 findings  \\ \hline
\textbf{multiple\_entrypoint\_2} & 4 tainted flows to 3 sinks  & 5 flows, 3 sinks and 7 findings  \\ \hline
\textbf{multiple\_leakage\_1} & 10 tainted flows to 4 sinks  & 7 flows, 5 sinks and 5 findings  \\ \hline
\textbf{multiple\_leakage\_2} & 6 tainted flows to 5 sinks  & 11 flows, 5 sinks and 9 findings \\ \hline
\textbf{multiple\_leakage\_3} & 5 tainted flows to 5 sinks  & 37 flows,   7 sinks and 16 finding  \\ \hline
\textbf{side\_channel\_1} & No leakage  & No leakage \\ \hline
\textbf{side\_channel\_2} & 3 tainted flows to 3 sinks & 6 flows, 3 sinks and 2 findings  \\ \hline
\end{tabular}
\end{adjustbox}
\vspace{-0.5 cm}
\end{table}

  \hlc[highlight]{The answer to our \textit{RQ1} is:} The core module of Taint-Things accurately detects the same tainted flows that SAINT detected while showing significant improvement in speed. Celik et al. \cite {Celik} reported on the SAINT's results on a 230 dataset, using a 2.6GHz 2-core Intel i5 processor and 8GB RAM took around 16 minutes to evaluate the batch, while an individual app took $23\pm5$ seconds on average. On the other hand, our tool achieved significant improvement in performance with at least 4 folds. In addition, our tool was able to analyse apps that SAINT times out or fails on startup. detailed results of performance analysis of our tool is presented in Table \ref{tab1:perform}. This improvement is mainly because our approach computes dependency chains directly form the code, using the inductive transformation paradigm, while SAINT builds a dependency graph for the program and then use graph algorithms to reduce it to the tainted flow, which is mapped back to source statements afterward. 
  \begin{table*}[!b]
\vspace{-0.3cm}
\centering
\begin{adjustbox}{width=1\textwidth,totalheight={1\textheight},keepaspectratio}%
\begin{tabular}{|l|l|l|l|l|}
\hline
\textbf{App Category} & \textbf{Including Warning-producing Apps} & \textbf{Excluding Warning-producing Apps} & \multicolumn{1}{c|}{\textbf{\begin{tabular}[c]{@{}c@{}}Timeout\textbackslash Warning\\ Taint-Things\end{tabular}}} & \multicolumn{1}{c|}{\textbf{\begin{tabular}[c]{@{}c@{}}Timeout\textbackslash Failed\\ SAINT\end{tabular}}}  \\ \hline
Community & 1m25s (Total apps: 143 ) & 0m54s  (Total apps: 138 ) & 6 & 24 \\ \hline
Forum & 2m57s  (Total apps: 56 ) & 0m25s  (Total apps: 48 ) & 11 & 12 \\ \hline
Market & 0m27s  (Total apps: 42 ) & 0m14s  (Total apps: 41 ) & 1 & 6 \\ \hline
Benchmarks & 0m9s  (Total apps: 19 ) & 0m9s  (Total apps: 19 ) & 0 & 0 \\ \hline
\textbf{Total} & \textbf{4m4s (Total apps: 260 )} & \textbf{1m26s (Total apps: 246)} & \textbf{18} & \textbf{42} \\ \hline
\end{tabular}
\end{adjustbox}
\vspace{0.2cm}
\caption{Performance analysis}
\label{tab1:perform}
\end{table*}

  \subsection{Evaluating Flow-sensitive Analysis}
    \hlc[highlight]{To answer \textit(RQ2), evaluate the results of our implementation with flow-sensitive analysis,} we compared the results of our tool with and without the application of SSA. We used the previous dataset, which includes 260 IoT apps gathered from different sources. For the sake of consistency we have excluded programs that gave errors in the final results; this included 4 programs without SSA, and another 4 when adding due to the SSA TXL transformation failing on some apps. As well as 35 apps using SAINT web app at the time of the dataset collection. Table \ref{tabl:SSAresults2} shows the results of this test. The criteria is based on the previous comparison done in the previous section. We see that Adding SSA does not affect the final result which shows that no false positives were caused from the lack of flow sensitivity. 
   \begin{table}[t!]
\caption{SSA test results}
\label{tabl:SSAresults2}
\begin{adjustbox}{width=.7\textwidth, keepaspectratio}
\begin{tabular}{|l|c|c|c|c|c|c|}
\hline
\textbf{} & \multicolumn{2}{l|}{\textbf{Taint-Things}} & \multicolumn{2}{l|}{\textbf{SSA}} & \multicolumn{2}{l|}{\textbf{SAINT}} \\ \hline
 & \multicolumn{1}{l|}{Benign} & \multicolumn{1}{l|}{Leaking} & \multicolumn{1}{l|}{Benign} & \multicolumn{1}{l|}{Leaking} & \multicolumn{1}{l|}{Benign} & \multicolumn{1}{l|}{Leaking} \\ \hline
\textbf{Community}   & 67 & 55 & 67 & 55 & 69 & 53 \\ \hline
\textbf{Forum}       & 16 & 28 & 16 & 28 & 17 & 27 \\ \hline
\textbf{Marketplace} & 11 & 24 & 11 & 24 & 13 & 22 \\ \hline
\textbf{Benchmark} & 01 & 18 & 01 & 18 & 01 & 18 \\ \hline
\end{tabular}
\end{adjustbox}
\end{table}
    \begin{table}[t!]
\caption{Detailed results when running Taint-Things with and without SSA form on the benchmarks set}
\begin{adjustbox}{height=0.9 \height, keepaspectratio}
 \label{tabl:SSABenchDet}
\begin{tabular}{|l|l|l|l|}
\hline
\textbf{} & \textbf{Taint-Things} & \textbf{Adding SSA} \\ \hline
\textbf{call\_by\_reflection\_1} & \begin{tabular}[c]{@{}l@{}}Warning followed by \\ 10 flows to 4 sinks \end{tabular}& \begin{tabular}[c]{@{}l@{}}Warning followed by \\   4 flows to 4 sinks\end{tabular}\\ \hline
\textbf{call\_by\_reflection\_2} & 2 flows to 2 sinks & 2 flows to 2 sinks \\ \hline
\textbf{call\_by\_reflection\_3} & 7 flows to 3 sinks & 7 flows to 3 sinks  \\ \hline
\textbf{explicit} & 6 flows to 3 sinks & 6 flows to 3 sinks   \\ \hline
\textbf{global\_variable\_1} & 7 flows to 3 sinks & 7 flows to 3 sinks \\ \hline
\textbf{implicit\_1} & 3 flows to 2 sinks & 3 flows to 2 sinks   \\ \hline
\textbf{implicit\_2} & 4 flows to 3 sinks & 4 flows to 3 sinks  \\ \hline
\textbf{implicit\_explicit} & 9 flows to 3 sinks & 9 flows to 3 sinks  \\ \hline
\textbf{leaking\_via\_closures} & 5 flows to 3 sinks & 5 flows to 3 sinks \\ \hline
\textbf{multiple\_devices\_1} & 2 flows to 2 sinks & 2 flows to 2 sinks  \\ \hline
\textbf{multiple\_devices\_2} & 2 flows to 1 sink & 2 flows to 1 sink  \\ \hline
\textbf{multiple\_devices\_3} & 2 flows to 1 sink & 2 flows to 1 sink \\ \hline
\textbf{multiple\_entrypoint\_1} & 11 flows to 3 sinks & 11 flows to 3 sinks \\ \hline
\textbf{multiple\_entrypoint\_2} & 4 flows to 3 sinks & 4 flows to 3 sinks \\ \hline
\textbf{multiple\_leakage\_1} & 10 flows to 4 sinks & 10 flows to 4 sinks \\ \hline
\textbf{multiple\_leakage\_2} & 6 flows to 5 sinks & 6 flows to 5 sinks  \\ \hline
\textbf{multiple\_leakage\_3} & 5 flows to 5 sinks & 5 flows to 5 sinks  \\ \hline
\textbf{side\_channel\_1} & No leakage & No leakage \\ \hline
\textbf{side\_channel\_2} & 3 flows to 3 sinks & 3 flows to 3 sinks  \\ \hline
\end{tabular}
\end{adjustbox}
\vspace{-0.5 cm}
\end{table}

  When looking in detail at the results of our tool when adding SSA form on this dataset and comparing them to the original results, we saw that the results are identical with few exceptions. This shows us that on a real dataset, most programs do not get affected by adding flow sensitive analyses and there aren't many cases of variables reassignment in a way that changes the dataflow. We have included a detailed look at the results on the benchmark apps from the dataset in Table \ref{tabl:SSABenchDet}. We can see that the results are identical with the exception of \textit{call\_by\_reflection\_1} where the SSA form actually helps in reducing the reported flows; in this case, without SSA a variable was being conflated with a possible source and thus flows from it were reported, while the SSA assigns the variable a unique name avoiding this mixup. 
   \begin{table}[b!]
\begin{center}
\caption{Runtime performance when using SSA}
\begin{adjustbox}{width=0.4 \textwidth, keepaspectratio}
\label{tabl:SSAruntime}
\begin{tabular}{|l|l|l|}
\hline
\textbf{} & \textbf{Without SSA} & \textbf{With SSA} \\ \hline
\textbf{Real time} & 6m3.505s & 4m14.684s \\ \hline
\textbf{User time}   & 0m1.745s & 0m5.365s \\ \hline
\textbf{System time} & 0m6.717s & 0m17.959s \\ \hline
\end{tabular}
\end{adjustbox}
\end{center}
\end{table}

  Performance wise, this functionality creates relatively little overhead. To measure that, we compare the time Taint-Things takes to run on the dataset, once with the SSA transformation chained to it and another without it. Table \ref{tabl:SSAruntime} shows that the SSA transformation had an increase in CPU time but this translated to having no significant effect on the real time.

  
    \hlc[highlight]{Since we want to provide a definitive answer to \textit{RQ2}, we want to accurately measure precision and recall.} To achieve that, we used an independently generated dataset through a mutation framework that was developed by Alalfi et al. \cite{Mutation, alalfi2021mutation}. The framework generates mutants targeting the evaluation of flow analysis by altering the statements' order in benign flows to generate leaking ones. The mutants covered multiple patterns where variable reassignment happen to make the flow tainted or benign and to be leaked in multiple sink types such as leakage through the internet, notifications and messages. To compute precision and recall we used the following two equations:   
  \[precision=\frac{True Positives}{True Positives+False Positives}\]
  \[recall=\frac{True Positives}{True Positives+False Negatives}\]

    
  This dataset included 526 mutants, 263 benign and 263 leaking, adding SSA makes it possible to accurately report benign apps, while without it, it will be falsely reported as leaky. On the full dataset, Taint-Things achieves 100\% recall with all the 263 leaking apps detected and 50\% precision with 263 true positives out of the 526 reported leaking apps, but when adding SSA it achieves 100\% recall and 100\% precision with 263 true positives out of the 263 apps. 
   \begin{table}[t!]
\caption{SSA test results}
\label{tabl:SSAresults}
\begin{adjustbox}{width=.7\textwidth, keepaspectratio}
\begin{tabular}{|l|c|c|c|c|c|c|}
\hline
\textbf{} & \multicolumn{2}{l|}{\textbf{Taint-Things}} & \multicolumn{2}{l|}{\textbf{SSA}} & \multicolumn{2}{l|}{\textbf{SAINT}} \\ \hline
 & \multicolumn{1}{l|}{Benign} & \multicolumn{1}{l|}{Leaking} & \multicolumn{1}{l|}{Benign} & \multicolumn{1}{l|}{Leaking} & \multicolumn{1}{l|}{Benign} & \multicolumn{1}{l|}{Leaking} \\ \hline
\textbf{Community}   & 67 & 55 & 67 & 55 & 69 & 53 \\ \hline
\textbf{Forum}       & 16 & 28 & 16 & 28 & 17 & 27 \\ \hline
\textbf{Marketplace} & 11 & 24 & 11 & 24 & 13 & 22 \\ \hline
\textbf{Benchmark} & 01 & 18 & 01 & 18 & 01 & 18 \\ \hline
\end{tabular}
\end{adjustbox}
\end{table}

  Overall we can see that adding flow sensitivity though SSA is inexpensive and doesn't create an overhead. While the real dataset didn't have cases where adding flow sensitivity makes a significant effect, this functionality can still help in avoiding the false positive case where a variable can be reassigned a a benign value and cases where variables can be conflated. This can be very useful if we apply mitigation to the dataflow, where we want it to be correctly marked as benign when applied.
   \begin{table*}[!t]
\centering
\caption{Flow sensitivity mutation test}
\begin{adjustbox}{width=0.7\textwidth,totalheight={0.8\textheight},keepaspectratio}
\begin{tabular}{|l|l|l|l|}
\hline
\textbf{}  &  \textbf{\#Apps} & \textbf{Without SSA} & \textbf{With SSA}  \\ \hline

\textbf{Leaking} & 263 & \makecell{263 Detected as Tainted \\ (True positive)} &  \makecell{263 Detected as Tainted \\ (True positive)}
 \\ \hline
 
\textbf{Benign} & 263 &  \makecell{263 Detected as Tainted \\ (False positive)} &  \makecell{263 Accurately detected as benign \\ (True negative)} \\ \hline

\end{tabular}
\end{adjustbox}
\end{table*}

   \hlc[highlight]{One thing to note when transforming code to SSA form is the challenge of dealing with the scope. In groovy, variables are declared global by default. To make a local variable, it uses the \textit{def} keyword and then any reference to it would be local. Also variables declared locally in a method will still be accessed by block statements in that method. This makes it hard to generalize a static-analysis approach for dealing with scopes. In our SSA TXL transformation, we put precedence on the local scope by taking the structure of the inputted program into consideration. One problematic aspect of this approach is that some block statements, such as control flow statements can modify variables outside their scope. Flow sensitivity by itself ignores that, which can result into false negatives, as it either ignores variable modification inside these blocks, or in the case of a conditional statement, it'll only deal with one branch of execution. To deal with this problem path-sensitivity can be used.}
      \begin{table}
\caption{Optimized path sensitivity results}
\begin{adjustbox}{width=0.8\textwidth, keepaspectratio}
\label{tabl:pathSenRes}
\begin{tabular}{|l|c|c|c|c|c|c|c|c|}
\hline
 & \multicolumn{2}{l|}{\textbf{Taint-Things}} & \multicolumn{2}{l|}{\textbf{SSA}} & \multicolumn{2}{l|}{\textbf{Path Sensitivity}} & \multicolumn{2}{l|}{\textbf{SAINT}} \\ \hline
 & \multicolumn{1}{l|}{Benign} & \multicolumn{1}{l|}{Leaking} & \multicolumn{1}{l|}{Benign} & \multicolumn{1}{l|}{Leaking} & \multicolumn{1}{l|}{Benign} & \multicolumn{1}{l|}{Leaking} & \multicolumn{1}{l|}{Benign} & \multicolumn{1}{l|}{Leaking} \\ \hline
\textbf{Community}   & 63 & 47 & 63 & 47 & 63 & 47 & 65 & 45 \\ \hline
\textbf{Forum} & 15 & 21 & 15 & 21 & 15 & 21 & 13 & 23 \\ \hline
\textbf{Marketplace} & 11 & 21 & 11 & 21 & 11 & 21 & 11 & 21 \\ \hline
\textbf{Benchmark}   & 01 & 16 & 01 & 16 & 01 & 16 & 01 & 16 \\ \hline
\end{tabular}
\end{adjustbox}
\end{table}
  \subsection{Evaluating Path-sensitive Analysis}
  \hlc[highlight]{To answer \textit{RQ2} and evaluate our approach for path sensitivity,} we have run the program on the original dataset. We have gathered the results and compared them to the previous findings of our tool and SAINT's.
  \begin{table}
\begin{center}
\caption{Detailed results of optimized path generation and analyses for the benchmarks Set}
\begin{adjustbox}{width=0.7\textwidth, keepaspectratio}
\label{tabl:PathBenchDet}
\begin{tabular}{|l|l|l|l|l|l|}
\hline
 & \textbf{\#Paths} & \textbf{Benign Paths} & \textbf{Leaky Paths} & \textbf{\#Flows}\\ \hline
\textbf{call\_by\_reflection\_1} & Error & Error & Error & Error\\ \hline
\textbf{call\_by\_reflection\_2} & 12 & 10 & 2 (2 unique) & 2 \\ \hline
\textbf{call\_by\_reflection\_3} & 3 & 0 & 3 (2 unique) & 7 \\ \hline
\textbf{explicit} & 9 & 2 & 7 (5 unique) & 5\\ \hline
\textbf{global\_variable\_1} & 2 & 0 & 2 (2 unique) & 4\\ \hline
\textbf{implicit\_1} & 2 & 0 & 2 (2 unique) &  3 \\ \hline
\textbf{implicit\_2} & 3  & 0 & 3 (2 unique) & 4\\ \hline
\textbf{implicit\_explicit} & 3 & 0 & 3 (2 unique) & 7\\ \hline
\textbf{leaking\_via\_closures}  & 2  & 0 & 2 (2 unique) & 4\\ \hline
\textbf{multiple\_devices\_1} & 2 & 1 & 1 (1 unique) & 1\\ \hline
\textbf{multiple\_devices\_2} & 2 & 1 & 1 (1 unique) & 2 \\ \hline
\textbf{multiple\_devices\_3} & 3 & 0 & 3 (2 unique) & 2\\ \hline
\textbf{multiple\_entrypoint\_1} & 3 & 0 & 3 (2 unique) & 9\\ \hline
\textbf{multiple\_entrypoint\_2} & 3 & 0 & 3 (2 unique) & 4\\ \hline
\textbf{multiple\_leakage\_1}    & 3 & 0 & 3 (3 unique) & 6 \\ \hline
\textbf{multiple\_leakage\_2}    & 3 & 0 & 3 (3 unique) & 5\\ \hline
\textbf{multiple\_leakage\_3}    & 3 & 0 & 3 (3 unique) & 4\\ \hline
\textbf{side\_channel\_1} & 1 & 1 & 0 & 0\\ \hline
\textbf{side\_channel\_2} & Error & Error & Error & Error\\ \hline
\end{tabular}
\end{adjustbox}
\end{center}
\end{table}

  Table \ref{tabl:pathSenRes} shows a comparison between the results generated by Taint-Things with and without path sensitivity and SAINT. We used the optimization method described earlier, but it should be noted that there were still few apps with more than 12 if-statements that we had to skip, since they end with a large number of paths that would exceed the ability of our testing device to analyze in a practical manner. And for the sake of consistency we also skipped the apps the introduced errors just like previous comparisons.
  Since the path tool actually generates multiple cases for each program, the criteria we followed for the classification was if at least one leaking path was detected, the program is flagged leaking as such. Table \ref{tabl:pathSenRes} shows our findings. We saw that on a real dataset, there wasn't any miss-match related to path sensitivity when it comes to determining whether a program is benign or leaking. This shows that there were no false positives related to path sensitivity on the original test. For a more detailed look in the results, we've added the detailed results on the benchmark part of the set in Table \ref{tabl:PathBenchDet} where we examine how many paths were generated and how many of them are benign, denoting a possibility of running the program in a way that doesn't leak data. We also looked how many of the leaky paths contained unique new flows, as apposed to having the same flow reporting in each of them and finally the total number of flows reported in all paths. Generally the flows reported matches the ones in the SSA form previously, but the path sensitivity gives more detail to the cases where each can happen.
  
  In the original version of the path-sensitivity approach was also independently tested on a set of mutations that were based on the original dataset to include path-sensitivity related attacks. The set included 440 mutations with different leakages of different types; through messages, posting on the internet and notifications. 
  When tested, it was found that our tool managed to cover all the possible flows for all the mutations. From this set, our tool managed to catch all the tainted flows while avoiding false positives. The results on this set showed a high level of precision and recall; the precision is 100\% since there are no false positives reported in that set, and recall is calculated as 100\%.

 
 \subsubsection{Evaluating Context-sensitive Analysis}
  \hlc[highlight]{To answer \textit{RQ2} and evaluate our approach for context sensitivity,} we first wanted to make test case exemplifying the case and how it is handled by Taint-Things. Listing \ref{list:contextEvalExam} shows an example where \textit{message1} calls  \mbox{\textit{returnMehtod}} with a benign string and then gets sent through a sink. The expected result is that no tainted flow is to be reported. On SAINT, it correctly reports that there is no potential leakgage. But in Taint-Things we get a flow from the lines: 3 7 8 2 10. This shows a context insensitive behavior, where \textit{returnMehtod} got tagged as tainted because of \textit{message2}. And even though the context of  the method call in \textit{message1} is different, it still gets conflated and the flow is considered tainted because of that. But after we run the function cloning transformation on this example, Taint-Things doesn't report a tainted flow, which is the correct context-sensitive behaviour.
  
  \begin{lstlisting}[style=Groovy, float, floatplacement=H, linewidth= 0.9\textwidth, caption={Context Sensitivity Example },label=list:contextEvalExam ]
  def initialize () {
    message1 = returnMethod ("benign") 
    message2 = returnMethod ("$sensitiveData")
    sendSms (message1)
  }

  def returnMethod (x) {
    return x 
  }
  \end{lstlisting}
  
  We've also ran a test on the dataset to see how it affects the results. Like the previous comparisons, it didn't have a big difference on the dataset in terms of labeling apps as tainted or benign, thus showing that there were no false positives done due to context sensitivity in the dataset. We found that there were changes done on how many unique flows and sinks were detected as well as changes on the line numbers reported from the original source code. This is due to the flow running through the multiple newly made clone methods. Performance wise, the process of function cloning isn't heavy in itself, but as the program gets bigger and with more nested function calls, analyzing the results can be more costly.


  Additionally, we've tested the tool on a dataset of mutations that adds patterns requiring context sensitive analysis. With the exception of one mutation operator, our tool was able to detect all the tainted flows while avoiding false negatives. For that operator, due to its complexity, the tool had a problem parsing method calls and ended up conflating them. Overall it achieved 100\% precision and 96.8\% recall.

  \hlc[highlight]{ In the mutation testing, if we give equal weight to each of the three categories of mutators that address the three levels of sensitivity analysis, we can average their results to get an estimated overall precision and recall for each of the tools. When one file is counted as one mutant and the correctness of the path sensitivity results are considered SaINT has 100\% recall and 56.8\% precision, Taint-Things has 99\% recall and 100\% precision.
  
  For Taint-Things, it can distinguish the change from the created base file to the generated mutant. But it failed to identify the mutants generated from one app that contained an extensive usage of state variable which marked it aggressively as a potential source. It failed for all the benign equivalent mutants generated from one source app when we only had sixteen source apps. 
}

 \begin{table}[t!]
\caption{Detailed results when running Taint-Things with and without method cloning form on the benchmarks set}
\begin{adjustbox}{height=0.95 \height, keepaspectratio}
 \label{tabl:CloneBenchDet}
\begin{tabular}{|l|l|l|l|}
\hline
\textbf{} & \textbf{Taint-Things} & \textbf{Adding Cloning} \\ \hline
\textbf{call\_by\_reflection\_1} & \begin{tabular}[c]{@{}l@{}}Warning followed by \\ 10 flows to 4 sinks \end{tabular}& \begin{tabular}[c]{@{}l@{}}Warning followed by \\   23 flows to 10 sinks\end{tabular} \\ \hline
\textbf{call\_by\_reflection\_2} & 2 flows to 2 sinks & 4 flows to 2 sinks \\ \hline
\textbf{call\_by\_reflection\_3} & 7 flows to 3 sinks & 4 flows to 3 sinks \\ \hline
\textbf{explicit} & 6 flows to 3 sinks & 8 flows to 6 sinks \\ \hline
\textbf{global\_variable\_1} & 7 flows to 3 sinks & 17 flows to 7 sinks \\ \hline
\textbf{implicit\_1} & 3 flows to 2 sinks & 3 flows to 2 sinks \\ \hline
\textbf{implicit\_2} & 4 flows to 3 sinks & 6 flows to 6 sinks \\ \hline
\textbf{implicit\_explicit} & 9 flows to 3 sinks & 12 flows to 6 sinks \\ \hline
\textbf{leaking\_via\_closures} & 5 flows to 3 sinks & 12 flows to 6 sinks \\ \hline
\textbf{multiple\_devices\_1} & 2 flows to 2 sinks & 6 flows to 6 sinks \\ \hline
\textbf{multiple\_devices\_2} & 2 flows to 1 sink & 6 flows to 3 sink \\ \hline
\textbf{multiple\_devices\_3} & 2 flows to 1 sink & 2 flows to 2 sink \\ \hline
\textbf{multiple\_entrypoint\_1} & 11 flows to 3 sinks & 16 flows to 6 sinks \\ \hline
\textbf{multiple\_entrypoint\_2} & 4 flows to 3 sinks & 6 flows to 6 sinks \\ \hline
\textbf{multiple\_leakage\_1} & 10 flows to 4 sinks & 24 flows to 8 sinks \\ \hline
\textbf{multiple\_leakage\_2} & 6 flows to 5 sinks & 14 flows to 10 sinks \\ \hline
\textbf{multiple\_leakage\_3} & 5 flows to 5 sinks & 10 flows to 10 sinks \\ \hline
\textbf{side\_channel\_1} & No leakage & No leakage  \\ \hline
\textbf{side\_channel\_2} & 3 flows to 3 sinks & \begin{tabular}[c]{@{}l@{}}Warning followed by \\ 7 flows to 7 sinks\end{tabular} \\ \hline
\end{tabular}
\end{adjustbox}
\vspace{-0.5 cm}
\end{table}
     
\section{Related Work}\label{ch:Survey}
 
  IoT is still new technology, yet diverse, and it poses many security challenges. To get an overview of the field, we look into the previous research done on IoT security. Furthermore, to get a better understanding of the techniques and approaches, we have to look outside of IoT research and into different fields such as android apps security, which shares some similar features, but had more time to mature.
  
  \subsection{IoT Security}\label{sec:IoTSurvey}
  Since the field of program analysis for the IoT is still in its infancy, there is only few related work on this area. Fernandes et al. \cite{b1} presents an approach for exposing vulnerabilities in SmartTings IoT apps. They concluded that many of the existing applications have vulnerabilities, mainly in the form of over-privilege. This study opened the field for later research to investigate the security aspects from a program analysis point of view, trying to provide potential solutions or ways to detect these problems. 
  
  Tian et al. \cite{b2} proposed a semantic based approach, with the objective of better representing applications' functionality and privileges to users. While Wang et al. \cite{b3} dealt with the logging problem and interconnectivity. With attention on the privacy aspect of IoT security, Celik et al. \cite{Celik} tried to programmatically detect sensitive information used in apps where breaches might happen. Their research introduced SAINT, which is a static analysis tool that tries to detect tainted flow through IoT apps' code, which could lead to sensitive data leakage. SAINT uses Groovy AST API to help recover an intermediate representation (IR) where taint sinks and sources are identified. They proposed using IR as means of abstracting the code, focusing on the important parts which might make the analysis easier. Sensitive data flow is then detected and reported if it is a feasible flow; meaning, the code can execute the sink function and leak data through it. 
  
  Celik et al. \cite{b5} also used this approach and applied it on abuse prevention, the safety and security aspects of IoT apps. They introduced SOTERIA which performs static analysis check to find potential vulnerabilities in apps where it is tested against safety, security and functionality properties. This could be used both on single apps or in multi-app environments.
  
  We focus in our research on the issue of privacy leaks using static analyses, so we looked more into the literature dealing with IoT app analysis. A literature review by Celik et al. \cite{b7} in 2018, surveyed six available tools that does privacy and security analyses, one of them, Saint, does static analysis for data leaks detection. 
  
  This comparison looked at multiple features concerning IoT specific issues, handling of app idiosyncrasy and analysis sensitivity. Specific issues include: Multi-app analysis, trigger-action platform support, proactive defense, lack of runtime prompts. Idiosyncrasies include: RESTful APIs, Closures and calls by reflection. Analysis sensitivities includes flow, context, field, path, and provenance tracking.
  
  According to the report Saint provides analysis sensitivity for all types and handles the mentioned idiosyncrasies. As well as having no runtime prompts and providing proactive defense. The criteria for path and context sensitivity that was considered is that the tool does not run infeasible paths. Saint achieves that by pruning these paths in the IR using a work list approach. A more detailed criteria would be considered

  Similar to SAINT, we offer another take on the problem of detecting potential privacy leaks in the source code, with the goal of adding more efficiency and exploring ways to improve the precision. SAINT and SOTERIA  implement their algorithms on the AST of a SmartThings app because of the constraints on Groovy language and proprietary back-end libraries. However, in our approach we compute dependency chains directly from the SmartThings App's code, using an inductive transformation paradigm. Our experiments shows that our approach produces equivalent results with significant improvement in performance, in speed and memory usage.   

  \subsection{Static Analysis in Android Apps}\label{sec:AndroidSurvey}
  
  We wanted to look more in depth in applications of static analysis and how sensitivity can be applied to it, so we looked into tools that run on Android apps. One thing to note in Android apps is that they have a well-defined IR. So, analyses can be done directly on that. Soot\cite{soot} and WALA\cite{WALA} are tools commonly used to convert the source code to its IR. 
  
  This approach is what an IFDS framework \cite{IFDS} does; the code is considered as an inter-procedural, finite, distributive subset, where it can be represented as a collection of flow graphs, with the statements and procedures as nodes. The analyses, then, is treated a problem of graph reachability. So in this case, an IR can be used to construct call graphs and used for the analyses.

  We started by looking into FlowDroid \cite{flowdroid} which is the first static analysis tool which provides full context, field, object, and flow sensitivity. It tries to solve problems that were not handled in previous tools. They are coarse grained and sometimes over or under approximate. Such problems happen when the life-cycle is not faithfully modeled, so the tool misses flows.
  
  It is uses the Soot framework for representing the Java code and IR. It works by analyzes the byte-code and configuration file, making a dummy main method and constructing the call graph to emulate the life-cycle. It then preforms the analysis on the call graph.
  
  The analysis is on-demand, based on IFDS framework, which adds context sensitivity, and is inspired by another tool, Andromeda, but adds more precision. For example, Andromeda can sometimes lack flow sensitivity, whereas FlowDroid adds that by using activations statements. One thing to note is that the IFDS framework and FlowDroid in extension are not path sensitive. It instead joins analysis results immediately at any control-flow merge point. Adding path sensitivity considered expensive.

  In literature review of static analysis tools for android apps \cite{androidSurvey}, the authors looked in multiple tools and their precision. One of their findings is that path sensitivity was often overlooked, with only 5 out of 30 of the surveyed tools provided it: Woodpecker\cite{woodpecker}, Apparecium\cite{apparecium}, Anadroid\cite{anadroid}, THRESHER\cite{thresher} and ContentScope\cite{contentscope}.
  
  Apparecium detects arbitrary data flows. Avoiding entry point analysis, it directly uses the sinks and sources. It uses textual representation, smali, for the code which is used to generate the class hierarchy. It then uses backward slicing for variables that can be assigned to sinks, followed by forward slicing for variables that can contain a source, and then combine these. It uses a data flow graph for its representation and can add paths to it.
  
  Woodpecker detects capability leaks in apps, which are cases where an app gains permissions without requesting. Builds a control flow graph from the byte code. It adds refines it path sensitivity using symbolic path simulation. Anadroid detects malware in android apps. And uses higher-order pushdown analyses and entry point saturation. ContentScope specifically tries to detect two vulnerabilities, passive content leak and content pollution and uses the analyses to determine their prevalence in Android markets. For detection of passive content leaks, it generates call graphs and tests for reachability.  Thresher deals with the tries to detect heap reachability using static analysis.
  
  These studies show different cases where static analysis is used in phone apps, specifically in Android, to detect taint flows, capability leaks, specific vulnerabilities or arbitrary flows. While there is some similarities in the concepts used in Android apps and IoT apps, like both being event driven models, there are differences and challenges specific to IoT. \hlc[highlight]{One problem is that while Groovy gets compiled to Java byte code, using the Android approaches is challenging due to Groovy's dynamic nature which makes it difficult to perform binary analysis with tools such as Soot. Other differences include that} Android apps can have a straightforward IR, where IoT platforms on the other hand use different programming languages, each with its own features and quirks that should be taken in consideration in the analyses. Naturally, this makes approaches based on tools specifically tailored for Android, such as Soot and WALA, not applicable to IoT. And while SAINT proposed introducing an IR for SmartThings IoT apps for the analysis, we propose doing the analysis directly on the source code. This is because SmartThings apps are generally less complicated, usually smaller in size, contained in one file and have a simple structure, due to the use of a sandbox environment, where the programming language features are limited; for example you cannot define your own classes when writing SmartThings apps.
  
  \hlc[highlight]{ However, all the surveyed taint flow analysis techniques for Android did not provide path sensitivity analysis as it is an expensive analysis. While we do provide a light weight path sensitivity analysis without compromising performance. Our analysis approach can be adapted to other platforms such as Android. Other members from the team already adapted the core taint analyser to analyse android apps} \cite{9252027}. \hlc[highlight]{Their  analysis provided a more accurate with improvement to  performance when compared to FlowDroid} \cite{flowdroid}.\hlc[highlight]{ However, that approach is not yet expanded to enable flow-, path- and context- sensitivity analysis.}

\section{Discussion}\label{ch:Discussion}
  
  In this paper We present a tainted flow static analysis approach for the identification and reporting of information leakage in Smarthings IoT apps.\hlc[highlight]{ For RQ1, we show that implementing the core analysis directly on the source code with less preprocessing can achieve 4 fold the speed while achieving the same results as the available tools. This is possible due to the simple and defined structure of SmartThings app.} 
 
\hlc[highlight]{Our approach automatically transform the source into SSA form to make the analysis flow sensitive. This approach avoids false positives that happen due to conflation of variables in the case of reassigning their values. We provide a framework for path generation to make the analysis path sensitive and for it to consider the branching in code execution and explored ways of optimizing the process to make it more applicable. We also explored function cloning as a way to make the analysis context sensitive and to avoid the conflation of method calls.

When compared to SaINT} \cite{saint} \hlc[highlight]{which parses SmartThings apps, transforms them into an intermediate representation and then constructs CFG to reason about taint flow analysis. Our approach applies an inductive transformation paradigm on the abstract syntax tree (AST) of the original source code. The AST is produced as part of the source transformation stage and is very much influenced by the grammar definition and overrides. There is no four step process like Saint (Parse, transform to IR, build CFG, then perform taint flow analysis). Instead our taint analysis is applied on the original source codes' AST. This is true for the core analysis. Providing more sound analysis requires some pre-transformations, such as SSA form for flows-sensitivity analysis, methods cloning for context sensitivity analysis, and path exploration module for path sensitivity analysis.}

  \hlc[highlight]{Furthermore, for RQ2, we show that flow, path and context sensitivity can be added as extra modules before the core analysis and would improve the precision of the tool by avoiding false positives. We found that the real dataset didn't contain major cases that generated false positives due to the lack of sensitive analysis, but provide them as a way of increasing the precision and avoiding potential false positives especially in the cases where a mitigation is introduced.
  
  To confirm our findings, we reference an extensive study conducted by Alalfi et al.} \cite{Mutation} \hlc[highlight] {to evaluate taint analysis tools for IoT applications using a mutation-based framework. The analysis evaluated Taint-Things with another two tools, SaINT and FlowsMiner. This study provides a clearer assessment of the tools' accuracy. It also tests the  consistency of the tool's results over a large number of test cases.

In their flow-sensitive mutation tests they found that SaINT and Taint-Things have a recall rate of 100\% for this calculation, but SaINT's precision rate dropped to 50\% where Taint-Things' precision remains 100\%, indicating that SaINT was not able to avoid false positives that were due to low sensitivity. 
    
For testing the impact path-sensitivity mutators, they checked the correctness of the tool's reports. SaINT was able to identify the tainted results, giving a recall of 100\% but had its precision drop by ignoring the potential benign paths if another tainted path existed and reported a false positive in mutants that had two benign paths. Taint-Things on the other hand is able to give a detailed report of all the potential paths

When it comes to context-sensitivity mutators, the results confirmed that the tools are using context-sensitive analysis up to a certain level. For SaINT, even after getting the context, it failed to differentiate benign from malicious when using a specific sink.

If we give equal weight to each of the three categories of mutators that address the three levels of sensitivity analysis, we can average their results to get an estimated overall precision and recall for each of the tools. When one file is counted as one mutant and the correctness of the path sensitivity results are considered SaINT has 100\% recall and 56.8\% precision, Taint-Things has 99\% recall and 100\% precision.
For Taint-Things, it can distinguish the change from the created base file to the generated mutant. But it failed to identify the mutants generated from one app that contained an extensive usage of state variable which marked it aggressively as a potential source. It failed for all the benign equivalent mutants generated from one source app when we only had sixteen source apps.} 

  \hlc[highlight]{ In general, we see that SmartThings real cases are not complex. From our tests, we have seen that all the apps from our datasets were not affected by the inclusion of the sensitive analyses. That is because most of them are simple one page apps written in the SmartThings sandbox} \cite{Sandbox} \hlc[highlight]{. This also allows our approach to avoid dealing with class hierarchy beyond the syntax analyses done on the  TXL grammar side, since so users can not define their own classes. Tainted flows exists in the events and functions and our analysis focuses on tracking them in the source code. Similarly we do not deal with pointer analysis. But it should be noted, while real cases do not exhibit complexity that gets affected by the sensitivities, their inclusion provides an important addition not only to minimize false positives but also to avoid false negatives, especially in the case of mitigating leaking apps.}
\section{Conclusions}\label{ch:conclusions}

\hlc[highlight]{In this paper we present a tainted flow static analysis approach for the identification and reporting of informationleakage in Smarthings IoT apps. Our approach provides a quicker core analyzer as well as more accurate analysis by adding flow- path- and context sensitivity analysis as added modules. When compared to existing tools addressing the same problem, our analysis provides more accurate results with a considerable higher performance gain. One aspect that may have improved the performance over other existing techniques is that the approach performs the analysis directly on the source. In addition, our approach is unique in providing a lightweight path sensitivity analysis, and innovative way to implement context sensitivity analysis using methods cloning and source transformation.}  
        
  We have deployed a version of Taint-Things that accounts for flow sensitivity analysis online as well. Taint-Things is currently available, for testing, at: http://taint-things.scs.ryerson.ca/ .
  
\section{Future Work}\label{ch:FutureWork}
  Future work includes extending our analysis approach to other Smart Home platforms, such as OpenHab. OpenHab uses Java, which already have a TXL grammar \cite{JavaGrammar}, but it also utilizes a domain specific language for handling its rules. The syntax for this is shared with Xtend. We can use its documentation  \cite{Xtend} to build a TXL grammar to handle it. OpenHab also uses a different structure than SmartThings apps and will have different definitions for sources and sinks. A study of how the sources and sinks appear in it, as well as modification of Taint-Things' transformation rules will be  required to accommodate that.
  
  While our approach examines potential leakage in each app by itself, further study can be done on potential leaks happening through multiple apps when they communicate with each other. SmartThings apps are usually self contained in one file and TXL can handle that easily as an input, but extending the program to handle multiple files and analyze their inter connectivity can be challenging. Nonetheless we can explore similar methods to what were used in adding flow and path sensitivity analysis, where we either chain and redirect the analysis results and use a script to work on multiple inputs.
  
  More ways to optimize and utilize precision can be explored. When adding flow sensitivity analysis, different ways to determine the variable scopes and using them in making the SSA form. This can be done through TXL and would require a rewriting of the script we currently use by altering the priorities of how variables get parsed and renamed. Ways for making path-sensitivity analysis less intensive and for context-sensitivity analysis to be achieved while maintaining the structure and line numbers of the original source code should also be studied. A compromise approach can be used where instead of doing full path or context sensitivity, we can use a partial approach. This can be done similarly to the optimized path sensitivity, where we only provide the path and context sensitivity only for some method calls and if-statements. 
  
  The next step after detecting tainted flow would be offering ways of mitigation or suggestions to good coding patterns and practices. This would require studying and mapping mitigation methods to each type of leakage. We can replace non-secure patterns with secure ones using TXL. And finally, we can explore ways for deploying the app in a more user friendly manner such as linking it to an IDE.

\bigskip 








\bibliographystyle{ACM-Reference-Format}
\bibliography{references.bib}

\appendix
\clearpage
\section{Appendix} \label{appendix:a}
\hlc[highlight]{The following Listing} \ref{list:realExample} \hlc[highlight]{is a real example of a SmartThings app that examines a common tainted flow pattern through a sendPush sink. The sensitive information here is the location of the user and his presence. It should be noted that this tainted flow is not necessarily malicious, but it presents a potential pattern where sensitive information could leak, which requires more scrutiny in the review, in case the data is not sanitized or going through an insecure API.} 

 \begin{lstlisting}[style=Groovy, float=!hb, linewidth= 0.9\textwidth, caption={Real App Example},label=list:realExample, basicstyle=\tiny]
/**
 *  Copyright 2015 SmartThings
 *
 *  Licensed under the Apache License, Version 2.0 (the "License"); you may not use this file except
 *  in compliance with the License. You may obtain a copy of the License at:
 *
 *      http://www.apache.org/licenses/LICENSE-2.0
 *
 *  Unless required by applicable law or agreed to in writing, software distributed under the License is distributed
 *  on an "AS IS" BASIS, WITHOUT WARRANTIES OR CONDITIONS OF ANY KIND, either express or implied. See the License
 *  for the specific language governing permissions and limitations under the License.
 *
 *  Unlock It When I Arrive
 *
 *  Author: SmartThings
 *  Date: 2013-02-11
 */

definition(
    name: "Unlock It When I Arrive",
    namespace: "smartthings",
    author: "SmartThings",
    description: "Unlocks the door when you arrive at your location.",
    category: "Safety & Security",
    iconUrl: "https://s3.amazonaws.com/smartapp-icons/Convenience/Cat-Convenience.png",
    iconX2Url: "https://s3.amazonaws.com/smartapp-icons/Convenience/Cat-Convenience%402x.png",
    oauth: true
)

preferences {
    section("When I arrive..."){
        input "presence1", "capability.presenceSensor", title: "Who?", multiple: true
    }
    section("Unlock the lock..."){
        input "lock1", "capability.lock", multiple: true
    }
}

def installed()
{
    subscribe(presence1, "presence.present", presence)
}

def updated()
{
    unsubscribe()
    subscribe(presence1, "presence.present", presence)
}

def presence(evt)
{
    def anyLocked = lock1.count{it.currentLock == "unlocked"} != lock1.size()
    if (anyLocked) {
        sendPush "Unlocked door due to arrival of $evt.displayName"
        lock1.unlock()
    }
}

 \end{lstlisting}
\clearpage

\hlc[highlight]{The following listings show the output of the analysis on the provided app. Listing} \ref{list:realExampleMark} \hlc[highlight]{shows the sink marking step. In this case the line containing the sendPush is marked as such. Listing} \ref{list:realExampleBack} \hlc[highlight]{shows the backward tracing step, where the variable passed to the sink are traced through the program, line containing the variables are tagged with the line numbers they pass the variable to. Listing} \ref{list:realExampleResult} \hlc[highlight]{shows the final results when the app is run through the analyzers. The  line numbers where the flows exists are printing with the relevant lines are marked.}

\begin{lstlisting}[style=Groovy, float=hb, linewidth= 0.9\textwidth, caption={Real App Example Mark Sinks Step},label=list:realExampleMark ]
 preferences {
    section (" When I arrive...") {
        input " presence1 ", " capability.presenceSensor ", title : " Who ? ", multiple : true
    }
    section (" Unlock the lock...") {
        input " lock1 ", " capability.lock ", multiple : true
    }
}

def presence (evt)
{
    if (anyLocked) {
        < sink > sendPush " Unlocked door due to arrival of $ evt.displayName " < / > 
    }
}    
\end{lstlisting}
 

 \begin{lstlisting}[style=Groovy, float=hb, linewidth= 0.9\textwidth, caption={Real App Example Trace Backward Step},label=list:realExampleBack ]
preferences {
    section (" When I arrive...") {
        input " presence1 ", " capability.presenceSensor ", title : " Who ? ", multiple : true
    }
    section (" Unlock the lock...") {
        input " lock1 ", " capability.lock ", multiple : true
    }
}

def installed ()
{
    < 50 source > subscribe (presence1, " presence.present ", presence) < / >
}

def updated ()
{
    < 50 source > subscribe (presence1, " presence.present ", presence) < / >
}

def presence (< > evt < / >)
{
    if (anyLocked) {
        < sink > sendPush " Unlocked door due to arrival of $ evt.displayName " < / >
    }
}
\end{lstlisting}

\begin{lstlisting}[style=Groovy, float=hb, linewidth= 0.9\textwidth, caption={Real App Example Final Result},label=list:realExampleResult ]
41 50 54
47 50 54



preferences {
    section (" When I arrive...") {
        input " presence1 ", " capability.presenceSensor ", title : " Who ? ", multiple : true
    }
    section (" Unlock the lock...") {
        input " lock1 ", " capability.lock ", multiple : true
    }
}

def installed ()
{
    < 50 source > subscribe (presence1, " presence.present ", presence) < / >
}

def updated ()
{
    < 50 source > subscribe (presence1, " presence.present ", presence) < / >
}

def presence (< 54 > evt < / >)
{
    if (anyLocked) {
        < sink > sendPush " Unlocked door due to arrival of $ evt.displayName " < / >
    }
}

\end{lstlisting}
\end{document}